\documentclass[twocolumn, pra, superscriptaddress,reprint]{revtex4-1}
\usepackage[usenames,dvipsnames]{color} 
\usepackage{graphicx}
\usepackage{CJK}
\definecolor{LinkColor}{rgb}{0,0,.5}
\usepackage[colorlinks=true,linkcolor=BrickRed,citecolor=blue,urlcolor=LinkColor]{hyperref}

\usepackage{physics}
\newcommand{\ave}[1]{\left\langle #1\right\rangle}
\renewcommand{\tr}[1]{\textrm{Tr}\left[{#1}\right]}
\newcommand{\ve}{\boldsymbol}

\begin{document}
\begin{CJK*}{UTF8}{}

\title{A geometric perspective: experimental evaluation of the quantum Cram\'er-Rao bound }
\author{Changhao Li \CJKfamily{gbsn}(李长昊)}
\affiliation{
Research Laboratory of Electronics, Massachusetts Institute of Technology, Cambridge, Massachusetts 02139, USA
}
\affiliation{
Department of Nuclear Science and Engineering, Massachusetts Institute of Technology, Cambridge, Massachusetts 02139, USA
}

\author{Mo Chen \CJKfamily{gbsn}(陈墨)}
\affiliation{
Research Laboratory of Electronics, Massachusetts Institute of Technology, Cambridge, Massachusetts 02139, USA
}
\affiliation{
Institute for Quantum Information and Matter, California Institute of Technology, Pasadena, CA 91125, USA
}
\affiliation{
Thomas J. Watson, Sr., Laboratory of Applied Physics, California Institute of Technology, Pasadena, CA 91125, USA
}
\author{Paola Cappellaro}
\thanks{pcappell@mit.edu}
\affiliation{
Research Laboratory of Electronics, Massachusetts Institute of Technology, Cambridge, Massachusetts 02139, USA
}
\affiliation{
Department of Nuclear Science and Engineering, Massachusetts Institute of Technology, Cambridge, Massachusetts 02139, USA
}
\affiliation{
Department of Physics, Massachusetts Institute of Technology, Cambridge, Massachusetts 02139, USA
}

\begin{abstract}
The power of quantum sensing rests on its ultimate precision limit, quantified by  the quantum Cram\'er-Rao bound (QCRB), which can surpass classical bounds. 
In multi-parameter estimation, the QCRB is not always saturated, as the quantum  nature of associated observables may lead to their incompatibility. Here, we explore the precision limits of multi-parameter estimation through the lens of quantum geometry, enabling the experimental evaluation of the QCRB via quantum geometry measurements.  Focusing on two- and three-parameter estimation, we elucidate how fundamental quantum uncertainty principles prevent the saturation of the bound. 
By linking a metric of ``quantumness'' to the system's geometric properties, we investigate and
experimentally extract the attainable QCRB for three-parameter estimations.
\end{abstract}
\maketitle
\end{CJK*}

\section{Introduction}
Quantum sensing stands as a key application for quantum technologies~\cite{RevModPhys.89.035002}. 
Quantum sensors  promise to achieve better sensitivity or precision than classical systems and have been utilized in many fields, ranging from material science~\cite{Estefani2022,Thiel2019,Yip2019,Lesik2019,Hsieh2019} to biology~\cite{Kucsko2013,Choi2020,Li2022}. 
Quantum metrology quantifies the precision limit  of quantum sensing with the quantum Cram\'er-Rao bound (QCRB)~\cite{Helstrom1967,Yuen1973,Belavkin1976,Holevo1977,PARIS2009,Carollo_2019,Liu2019,Albarelli2020}. 
For unbiased estimation of an unknown system parameter, the QCRB is given by the inverse of the quantum Fisher information~\cite{Holevo1977,Carollo_2019,Liu2019,Albarelli2020}. The estimator is chosen to be optimal, therefore the QCRB only depends on the quantum state.  
In turn, the precision of estimating a parameter can be linked to the ``distance" between two nearby states differing by an infinitesimally small parameter change~\cite{BraunsteinPRL1994,GuoPRA2016,KolodrubetzPR2017,Liu2019}. This notion naturally connects quantum sensing to the quantum geometric properties of the system, as the quantum metric tensor is closely related to the ultimate estimation precision  quantified by the quantum Fisher information.

This picture becomes more involved when extending the goal from estimating a single parameter to multiple parameters, in analogy to the complexity of multi-dimensional systems versus 1D systems. The complication arises from the incompatibility between each parameter's optimal estimators--a signature of quantum mechanics~\cite{Albarelli2020,Liu2019,GuoPRA2016}. This leads to trade-offs among the estimation precisions of different parameters when picking the initial state. The interplay among observables makes the multi-parameter estimation problem more intriguing than  single-parameter estimation. Hence, developing tools to investigate the ultimate precision and the attainable QCRB in a multi-parameter estimation setting is of great interest for quantum metrology and sensing applications. 
A promising strategy, similar to the single-parameter scenario, is to link quantum geometric properties of the (multi-)parametrized state to the estimation precision. Specifically, in addition to the link between the quantum metric tensor and the quantum Fisher information, we can link  the non-commutativity of optimal measurement operators for different parameters to the Uhlmann curvature~\cite{Carollo_2019,Liu2019,Albarelli2020}, as we explain below. Characterizing the geometry of the parametrized quantum states by measuring the quantum geometric tensor would thus help explore the problem of quantum multi-parameter estimation and evaluate the attainable QCRB from a geometric perspective.

In this work,  we explore the relation between quantum geometry and multi-parameter estimation to gain novel insight into the corresponding precision bound. 
In particular,  geometry and metrology metrics are  connected via a \textit{characterization number} $\gamma$~\cite{Carollo_2019}, which characterizes the ``quantumness" of the system: 
$\gamma$ is  upper bounded by the fundamental uncertainty relation and  it can be a signature of exotic topological structures in parameter space. Beyond its theoretical insight, this relation provides a strategy to experimentally evaluate quantum estimation bounds. 
We can thus provide an experimental demonstration of these results by focusing on a three-parameter, three-level pure state model, which is synthesized using a single nitrogen-vacancy (NV) center in diamond. 
Based on measuring spin-1 Rabi oscillations upon parameter-modulated driving, we develop experimental tools to extract the geometric quantities and determine the quantum Fisher information and Uhlmann curvature of the quantum state.
Then the system's quantumness, i.e., the incompatibility among parameters to be estimated, is characterized with the help of $\gamma$. We finally evaluate the attainable QCRB for the parameterized state, as well as the two-parameter states in the subspace of the three-parameter model.

\section{Relating quantum geometry to quantum multi-parameter sensing}\label{sec:Theory}

We consider a generic quantum multi-parameter estimation problem  given by a quantum statistic model $\rho_{\boldsymbol{\theta}}$, a family of density operators labeled by $\boldsymbol{\theta}=(\theta_1,\theta_2,...,\theta_d)^T$ that is the set of unknown parameters to be estimated. 
Performing POVM $\mathcal{M}=\{ \mathcal{M}_k | \sum_k \mathcal{M}_k =I, \mathcal{M}_k \geq 0 \}$, we  obtain the measurement conditional probabilities $p(k|\boldsymbol{\theta}) = \textrm{Tr}[\rho_{\boldsymbol{\theta}} \mathcal{M}_k]$. 
With $N$  repeated measurements, the unknown parameters are estimated via the estimator $\tilde{\boldsymbol{\theta}}(k)$. 
The  accuracy of the estimation is quantified by the mean-square error matrix $\boldsymbol{V(\boldsymbol{\theta})} = \sum_k p(k|\boldsymbol{\theta})  [\tilde{\boldsymbol{\theta}}(k)-\boldsymbol{\theta}]  [\tilde{\boldsymbol{\theta}}(k)-\boldsymbol{\theta}]^T $. 
Obtaining the lower bounds of this error matrix is of critical importance in quantum sensing and metrology. 

The bounds usually depend on the choice of POVM. 
The symmetric logarithmic derivative~\cite{Helstrom1967,Yuen1973,Belavkin1976,PARIS2009} (SLD) can be used to define  a bound that only depends on the quantum statistic model $\rho_{\boldsymbol{\theta}}$, having optimized over the measurement operators. 
The corresponding quantum Fisher information matrix (QFIM) then yields the matrix SLD QCRB $\boldsymbol{V(\boldsymbol{\theta})} \geq \boldsymbol{J}(\boldsymbol{\theta})^{-1} $. 
It is convenient and common to introduce a scalar QCRB, $C^S (\boldsymbol{\theta}, \boldsymbol{W})$ =$ \tr{\boldsymbol{W} \boldsymbol{J}(\boldsymbol{\theta})^{-1}}$. Here the  weight matrix $\boldsymbol{W}$, is a positive definite matrix that weighs the uncertainty cost of different parameters. The scalar QCRB inequality becomes $\tr{\boldsymbol{W} \boldsymbol{V}} \geq C^S (\boldsymbol{\theta}, \boldsymbol{W})$. 
We remark that the weight matrix is introduced to compare distinct experimental strategies specified by POVM, $\mathcal{M}$ (which affects $\boldsymbol{V}$), and  states $\rho_\theta$ (which determines QFIM), overcoming the challenge of  comparing matrices without appropriate normalization. Practically, the weight matrix can be introduced to put more weight on parameters of interest.
This SLD-CRB is generally not attainable due to the incompatibility of generators of different parameters. That is, the optimal measurement operators corresponding to different parameters do not commute with each other, making this scalar bound unreachable.  

Holevo derived~\cite{Holevo1977} a tighter scalar bound,  the Holevo Cram\'er-Rao bound (HCRB) $C^H (\boldsymbol{\theta},\boldsymbol{W})$, which is however only obtained as an optimization. 
More recently, it has been shown that these two scalar bounds  satisfy the following inequality~\cite{Carollo_2019}
\begin{equation}
\label{eq:inequality}
\begin{aligned}
     & C^S(\boldsymbol{\theta},\boldsymbol{W}) \leq  C^H(\boldsymbol{\theta},\boldsymbol{W})  \\ \leq
    & C^S(\boldsymbol{\theta},\boldsymbol{W}) + || \sqrt{\boldsymbol{W}} \boldsymbol{J}^{-1}\boldsymbol{F} \boldsymbol{J}^{-1} \sqrt{\boldsymbol{W}}||_1 \\ \leq 
    & (1+\gamma)C^S(\boldsymbol{\theta},\boldsymbol{W}) \leq   2C^S(\boldsymbol{\theta},\boldsymbol{W}),
\end{aligned}
\end{equation}
where $||\cdot ||_1$ denotes the trace norm and  $\boldsymbol{F}$ is known as the mean Uhlmann  curvature  with element  $F_{\mu \nu}=-\frac{i}{4}\textrm{Tr}[\rho_{\boldsymbol{\theta}} [L_{\mu},L_{\nu}]] $ (for pure states this reduces to the Berry curvature.)
The \textit{characterization number} $\gamma$ that appears in Eq.~(\ref{eq:inequality}) is  defined as
\begin{equation}\label{eq:gamma_def}
    \gamma = ||i 2\boldsymbol{J}^{-1} \boldsymbol{F} ||_{\infty},
\end{equation}
where $||\boldsymbol{A}||_{\infty}$ denotes the largest eigenvalue of  $\boldsymbol{A}$.   $\gamma$  characterizes the ``quantumness'' of the system, i.e., it quantifies the amount of incompatibility of the statistic model $\rho_{\boldsymbol{\theta}}$. As implicit in Eq.~(\ref{eq:inequality}) and explained hereafter, it can be  proved that $0 \leq \gamma \leq 1$~\cite{Carollo_2019}.

An explicit expression for an attainable scalar QCRB with weight matrix $\boldsymbol{W}$=$\boldsymbol{J}$, which monotonically increases with $\gamma$, was derived in Ref.~\cite{Matsumoto2000,Matsumoto_2002}
\begin{align}\label{eq:C_gamma_relation}
    C(\boldsymbol{\theta}) & = \tr{\textrm{Re}\left[\sqrt{\openone_d+ 2i \boldsymbol{J}(\boldsymbol{\theta})^{-1/2} \boldsymbol{F}(\boldsymbol{\theta}) \boldsymbol{J}(\boldsymbol{\theta})^{-1/2} }\right]^{-2}} \nonumber\\
    & = \sum_{i}\frac{2}{1+\sqrt{1-|\gamma_i|^2}},
\end{align}
where $\openone_d$ is the d-dimensional identity matrix and $\gamma_i$ is the i-th eigenvalue of $i 2\boldsymbol{J}^{-1} \boldsymbol{F}$.
The evaluation of this \textit{attainable QCRB} in terms of geometric quantities, and its  relation to SLD-CRB and HCRB, is  our key result.

We remark here that choosing $\boldsymbol{W}$=$\boldsymbol{J}$, i.e., amplifying the variance of parameters with larger QFI, thus balancing the precision of different parameters, leads to a simplified form of QCRB. More precisely, a particular choice of $\boldsymbol{W}$ corresponds to a unique change of parameterization. When $\boldsymbol{W}$=$\boldsymbol{J}$,  the QFI matrix under the new parameterization becomes $J^{'} = AJA^T = I$ where $A$ is defined by $W = A^{-1}(A^{-1})^T$ and $I$ is the identity operator. The transformation  corresponds to the LDL decomposition of the original QFI matrix. In this scenario, quantum geometry is Euclidean and the covariance of parameter generators vanishes, hence the estimators of different parameters are not correlated. Simultaneous measurement of all parameters is equivalent to individual measurement protocols in the asymptotic limit and one gets the same amount of information for each parameter.

\begin{figure}[bt]
\centering 
\includegraphics[width=0.48\textwidth]{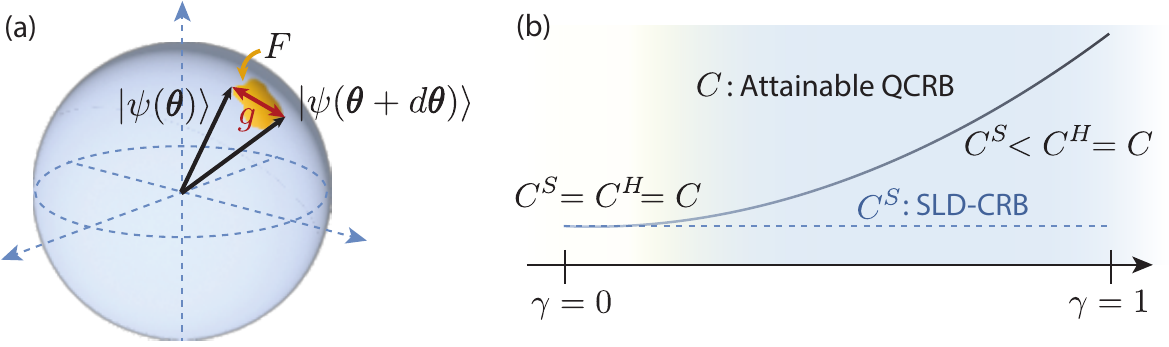}
\caption{\label{fig:fig1} \textbf{(a).} 
Diagram showing the ``distance'' between nearby states $\ket{\psi(\boldsymbol{\theta})}$ and $\ket{\psi(\boldsymbol{\theta} + d\boldsymbol{\theta})}$ in parameter space, from which the metric tensor $\boldsymbol{g}$ (proportional to quantum Fisher information $\boldsymbol{J}$) and Berry curvature $\boldsymbol{F}$ naturally arise.
\textbf{(b).} SLD-CRB and attainable QCRB for different $\gamma$ values. At $\gamma=0$, the model is quasi-classical and the SLD-CRB is reachable, i.e., $C^S=C^H=C$, while the model is called coherent at $\gamma=1$ with $C^S<C^H=C$. Here we take the weight matrix $\boldsymbol{W}=\boldsymbol{J}$  and denote $C^{S (H)}(\boldsymbol{\theta},\boldsymbol{W})$ as $C^{S(H)}$ for simplicity.
}
\end{figure}

To highlight the geometric interpretation of $\boldsymbol{J}, \boldsymbol{F}$ (so-far introduced via the SLD) and thus the connection between estimation and geometry, in the following we consider pure state models, i.e., $\rho_{\boldsymbol{\theta}} =\ket{\psi(\boldsymbol{\theta})}\! \bra{\psi(\boldsymbol{\theta})}$. 
 The geometric properties of this state are captured by the quantum geometric tensor (QGT), which naturally appears when one considers the ``distance" between nearby states $\ket{\psi(\boldsymbol{\theta})}$ and $\ket{\psi(\boldsymbol{\theta} + d\boldsymbol{\theta})}$ (Fig.~\ref{fig:fig1}): $
     ds^2 \equiv 1 - |\bra{\psi(\boldsymbol{\theta})} {\psi(\boldsymbol{\theta}\rangle + d\boldsymbol{\theta})}  |^2 = d\theta_\mu \chi_{\mu\nu} d\theta_\nu + O(|d\boldsymbol{\theta}|^3).$
 
Here we defined the QGT,  given by $\chi_{\mu\nu} = \bra{\partial_\mu \psi} (I-\ket{\psi}\bra{\psi})\ket{\partial_\nu\psi} = g_{\mu\nu} + \frac{i}{2}F_{\mu\nu}$. Its imaginary (antisymmetric) part $F_{\mu\nu}$ is the Berry curvature. The symmetric real part $g_{\mu\nu}$ 
is called the Fubini-Study metric tensor for the pure state case discussed here (also known as Bures metric for general mixed states) and characterizes the overlap between the nearby states. 
A larger $g_{\mu\nu}$ indicates a larger distinguishability between  two states differing by an infinitesimal change of parameters, and it is thus related to a larger quantum Fisher information. The role of the Berry curvature is subtler, as it links the attainability of the QFI bound in multi-parameter scenarios to the observable compatibility.

To show the relation between QGT and parameter estimation, one can assume the parameter $\boldsymbol{\theta}$ to be  introduced by evolving an initial state $\ket{\psi_0}$ via a unitary evolution, i.e., $\ket{\psi(\boldsymbol{\theta})} =U(\boldsymbol{\theta}) \ket{\psi_0} $. Then, the generator  for parameter $\mu \in \boldsymbol{\theta}$ is the operator  $\mathcal{G}_\mu \equiv i(\partial_{\mu}U) U^{\dagger}$. It can be shown (see Appendix~\ref{sec:AppendixQGT}) that the QFIM $J_{\mu\nu}$ and the metric tensor $g_{\mu\nu}$ are related by 
\begin{equation} \label{eq: relation_J_generator}
    J_{\mu\mu}\!=\!4 \ave{\Delta \mathcal G_\mu ^2}\!=\!4 g_{\mu\mu}, \quad
    J_{\mu\nu} \!=\! 4 \mathrm{Cov}(\mathcal G_\mu, \mathcal G_\nu) \!=\! 4 g_{\mu\nu},
\end{equation}
where $\ave{\Delta \mathcal G_\mu ^2}$ is the generator variance and $\mathrm{Cov}(\mathcal G_\mu, \mathcal G_\nu)$  their covariance.
Similarly, the Berry curvature is given by  the commutator between the parameter generators
\begin{equation}\label{eq: relation_F_generator}
    F_{\mu\nu} = i \ave{ [\mathcal G_\mu, \mathcal G_\nu ]}.
\end{equation}
It is now clear that the Berry curvature matrix characterizes the incompatibility of the system as it arises from the non-commutativity between generators of different parameters. 
When $F_{\mu\nu}=0$ ($\forall \mu,\nu \in \boldsymbol{\theta}$), the characterization number is $\gamma = 0$ according to Eq.~\ref{eq:gamma_def} and the system is quasi-classical. The discrepancy between the attainable QCRB and the SLD-CRB now vanishes, i.e., $C(\boldsymbol{\theta})=C^S(\boldsymbol{\theta},\boldsymbol{J})=C^H(\boldsymbol{\theta},\boldsymbol{J})=\tr{\openone_d}=d$.

As one of the main contributions of this letter, we next show that in the opposite limit the characterization number $\gamma$ is upper bounded by the fundamental quantum uncertainty principles, in contrast with the usual methods such as the proof in ~\cite{Carollo_2019}.
We consider the two-parameter case and start from the Robertson-Schr\"odinger uncertainty relation for two operators $\hat A, \hat B$~\cite{Robertson1934}, the stronger version of the better-known Heisenberg uncertainty relation (see also Appendix~\ref{sec:AppendixUncertainty}.)
In the quantum parameter estimation case, the operators $\hat A, \hat B$ can be replaced by the generators of parameters $\mu,\nu$, giving
\begin{equation} \label{eq: 2_generators_inequality}
\begin{split}
    \ave{\Delta \mathcal G_\mu^2} \ave{\Delta \mathcal G_\nu^2} \geq 
\frac{1}{4} \vert \ave{[ \mathcal G_\mu,  \mathcal G_\nu ]} \vert ^2 + \mathrm{Cov}(\mathcal G_\mu, \mathcal G_\nu)^2.
\end{split}
\end{equation}
With Eq.~(\ref{eq:gamma_def}), (\ref{eq: relation_J_generator}), (\ref{eq: relation_F_generator}) we have 
\begin{equation}
    \gamma = 2\frac{\sqrt{\mathrm{det}(\boldsymbol{F})}}{\sqrt{\mathrm{det}(\boldsymbol{J})}} = \frac{1}{2}\frac{\sqrt{\mathrm{det}(\boldsymbol{F})}}{\sqrt{\mathrm{det}(\boldsymbol{g})}} \leq 1,
\end{equation}
which relates $\gamma$ to  half  the  Berry's  phase along the curve that encloses a unit area induced by metric tensor. This relation holds for any two-parameter space in a multi-parameter setting.
The above relation can also be derived from the fact that the QGT $\chi_{\mu\nu}$ is a positive semi-definite matrix and has been explored in various models~\cite{Ozawa_BrunoPRB2021,Bruno_OzawaPRB2021,Bruno2022}.
Conversely, generalizing the two-operator uncertainty relations Eq.~\ref{eq: 2_generators_inequality} to multi-operators can be achieved using the fact that $\gamma\leq 1$, as we show in the Appendix~\ref{sec:Appendix}.

As a concrete example,  we consider a two-parameter model in a qubit system  $\ket{\psi(\theta, \phi)}=\cos\left(\frac{\theta}{2}\right) \ket0 -\sin\left(\frac{\theta}{2}\right)e^{-i\phi} \ket1$. The QFIM $\boldsymbol{J} =  \begin{pmatrix}
    1 & 0\\
    0 & \sin^2 \theta \\
    \end{pmatrix} $ and Berry curvature matrix  $
 \boldsymbol{F}= 
\begin{pmatrix}
    0 & \sin \theta/2\\
    -\sin \theta/2 & 0 \\
    \end{pmatrix}$
yield $\gamma =1 $, implying that the model is coherent and the two parameters $\theta, \phi$ here are  informationally exclusive. Indeed, they are associated with the non-commuting generators $\sigma_z$ and $\sigma_y$, respectively.

\begin{figure*}[htbp]
\centering 
\includegraphics[width=0.9\textwidth]{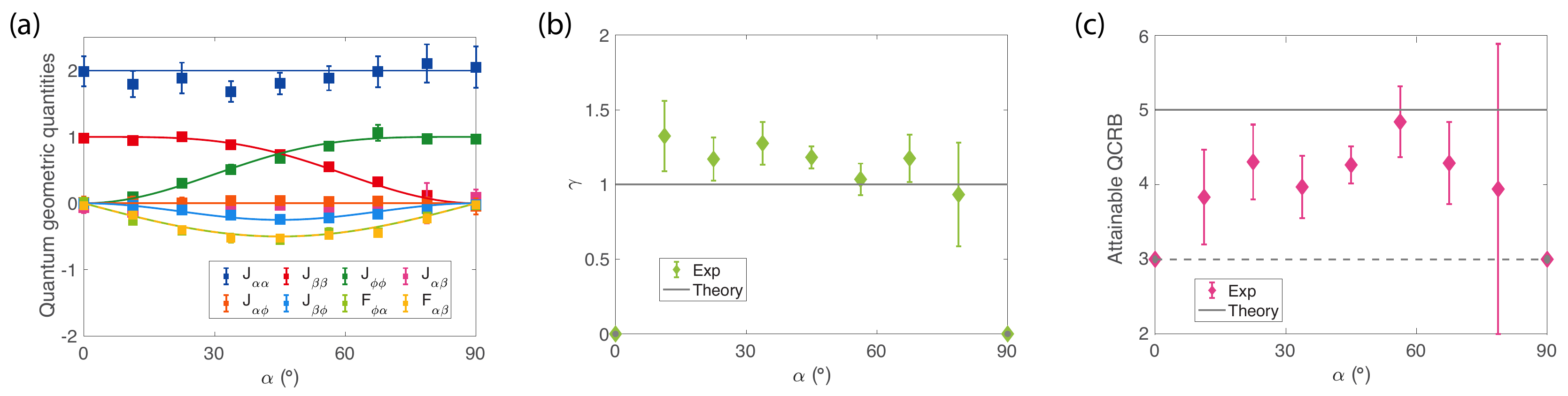}
\caption{\label{fig:fig2} \textbf{(a).} Experimental results on elements of quantum Fisher information matrix and Berry curvature matrix. The solid lines are theoretical calculations. \textbf{(b-c).} Characterization number $\gamma$ (b) and corresponding attainable QCRB (c) for the three-parameter model. The solid lines are theoretical predictions in both plots. The dashed line in (c) is the attainable QCRB when $\gamma=0$, i.e., the SLD-CRB for the model. The errorbars in (b-c) are propagated from the errors in quantum geometric quantities.} 
\end{figure*}

Eq.~(\ref{eq: relation_J_generator})-(\ref{eq: relation_F_generator}) show the connection between quantum geometry and quantum multi-parameter estimation (Fig.~\ref{fig:fig1}). The value of  $\gamma$ characterizes both the quantum multi-parameter estimation scenario, where it is a measure of the system's incompatibility, and  the geometry of the quantum state manifold, where it relates the  metric tensor determinant to the Berry curvature.  The ratio is indeed deeply connected to the existence of topological invariants that can characterize exotic monopole structures in parameter space. We leave the discussions to Appendix~\ref{sec:AppendixTopo}.

In the following, we experimentally explore the  relation between quantum geometry and quantum multi-parameter estimation. In particular, we measure the geometric quantities using a three-level system and demonstrate the relation~(\ref{eq:inequality})  obtaining the attainable QCRB. By embedding a two-parameter model into a 3-level system, we can further explore scenarios away from the two bounds $\gamma=0,1$.

\section{Experiments}
\label{sec:Experiments}
We engineer a pure state model using a single nitrogen-vacancy (NV) center in diamond at room temperature. The degeneracy of $\ket{m_s=\pm1}$ is lifted by an external magnetic field $B$=490 G along the NV quantization axis. This qutrit system can be fully controlled using dual-frequency microwave pulses  on-resonance with the transitions between $\ket{m_s=0}$ and $\ket{m_s=\pm 1}$. We can thus create any state parametrized by three angles,
\begin{equation} \label{eq:state_3para}
    \ket{\psi(\alpha,\beta,\phi)} = \frac{1}{\sqrt{2}}(cos(\alpha) e^{-i\beta}, -1,  \sin(\alpha) e^{-i\phi}  )^T.
\end{equation}
The state indeed corresponds to the ground state of a 3-band Hamiltonian that hosts exotic topological objects~\cite{chen2022}, see Appendix~\ref{sec:AppendixTopo}.
We measure the QGT of  this state to determine the QFIM and Berry curvature matrix  to characterize both its geometry and multi-parameter estimation properties. 
Specifically, we measure all the  components of the real  and imaginary part of the QGT separately using weak modulations of the parameters $\mu,\nu \in \{ \alpha, \beta,\phi \}$, a method that has been explored in previous work~\cite{chen2022,Ozawa2018,Yu2019}.

To extract the real part of the QGT, we first prepare the state in Eq.~(\ref{eq:state_3para}) with a two-tone microwave pulse after initialization to $\ket{m_s=0}$. 
We then apply driving with in-phase  modulation of the amplitudes $\mu, \nu$,  i.e., $\mu(t)=\mu_0 + m_\mu \sin (\omega t)$ and $\nu(t)=\nu_0 + m_\nu \sin (\omega t)$  with $|m_\mu|, |m_\nu| \ll 1$.  Here $\mu, \nu$ are the parameters in the state, i.e., $\mu, \nu \in \{ \alpha, \beta, \phi \}$.
Similarly, the imaginary part of the QGT can be extracted with out-of-phase modulations, where $\mu(t)=\mu_0 + m_\mu \sin (\omega t)$ and $\nu(t)=\nu_0 + m_\nu \cos(\omega t)$. 
These weak periodic modulations of the parameters will drive Rabi oscillations between the  state in Eq.~(\ref{eq:state_3para}) and the other two eigenstates of the qutrit system specified by Hamiltonian $H$, when the driving frequency $\omega$ is on-resonance with the $\Delta_1$ and $\Delta_2$, the energy splitting between $\ket{\psi(\alpha,\beta,\phi)}$ and the other two eigenstates $\ket{\psi_1} = (-\sin\alpha e^{-i\beta}, 0, \cos\alpha e^{-i\phi})^T$ and $\ket{\psi_2} = (\cos\alpha e^{-i\beta}, 1, \sin\alpha e^{-i\phi})^T$, respectively. 

The rates of these Rabi oscillations then correspond to the transition rates between states of interest Eq.~(\ref{eq:state_3para}) (ground state of $H$) and other eigenstates under perturbations. This captures the information encoded in the elements of QGT, which are coefficient of the second order term when defining the distance between nearby states. For example, to measure metric tensor component $g_{\alpha\alpha}$ at $(\alpha = \pi/4, \beta = 0, \phi = 0)$, we perform modulations in $\alpha$, i.e., $\alpha(t) = \pi/4 + m_\alpha \sin (\omega t)$ while fixing $\beta = 0, \phi = 0$. When setting $\omega = \Delta_1$ and $\omega = \Delta_2$, we observe coherent Rabi oscillations for the population remaining at  the initial state Eq.~\ref{eq:state_3para}  with rate $\Gamma_1$ and $\Gamma_2$, respectively. Then the metric tensor component is determined to be 
$g_{\alpha \alpha}=\frac{1}{4} J_{\alpha\alpha}= \frac{\Gamma_1^2}{\Delta_1^2}+\frac{\Gamma_2^2}{\Delta_2^2}$.
With this technique, by measuring the resulting Rabi frequencies (see Appendix~\ref{sec:AppendixExpt} for details),  we can extract the metric tensor matrix, and thus the QFIM, as well as the Berry curvature matrix.

The QGT components reconstructed using this technique are shown in Fig.~\ref{fig:fig2}(a). The measurement results are in good agreement with the theoretical predictions derived from the definition of quantum geometry tensor (details in Appendix~\ref{sec:AppendixQGT}).
The characterization number $\gamma$ can then be determined as in Fig.~\ref{fig:fig2}(b).
When $\alpha = 0$ or $\frac{\pi}{2}$, the parameterized model Eq.~\ref{eq:state_3para} effectively becomes a qubit system  in a one-parameter (sub)space. For example, when $\alpha =0$ we have $J_{\phi\phi}=0$, meaning that  we have no information on  $\phi$, and $\boldsymbol{F}=0$, indicating that there are no non-commuting generators, and the parameter $\beta$ can be estimated trivially. The model is thus quasi-classical. 
When $\alpha \in (0, \frac{\pi}{2})$, however, the model is coherent with $\gamma=1$, indicating the maximal non-commutativity of the generators  corresponding to the three parameters. 

We can further measure the attainable QCRB (Fig.~\ref{fig:fig2}(c)). Again, when $\alpha = 0$ or $\frac{\pi}{2}$, the attainable QCRB is equal to the SLD-CRB and HCRB, i.e., $C = C^S = C^H$ due to the quasi-classical nature of the system.
We note that small experimental errors  arising in separate QGT measurements are amplified by the relation Eq.~\ref{eq:C_gamma_relation}, leading to  a discrepancy of the attainable QCRB from theoretical predictions at $\alpha \in (0, \frac{\pi}{2})$. 
Nevertheless, our measurements reveal the key observation that when $\gamma=1$, the attainable QCRB is  
higher than SLD-CRB due to the incompatibility of the three parameters.

\begin{figure}
\centering 
\includegraphics[width=0.45\textwidth]{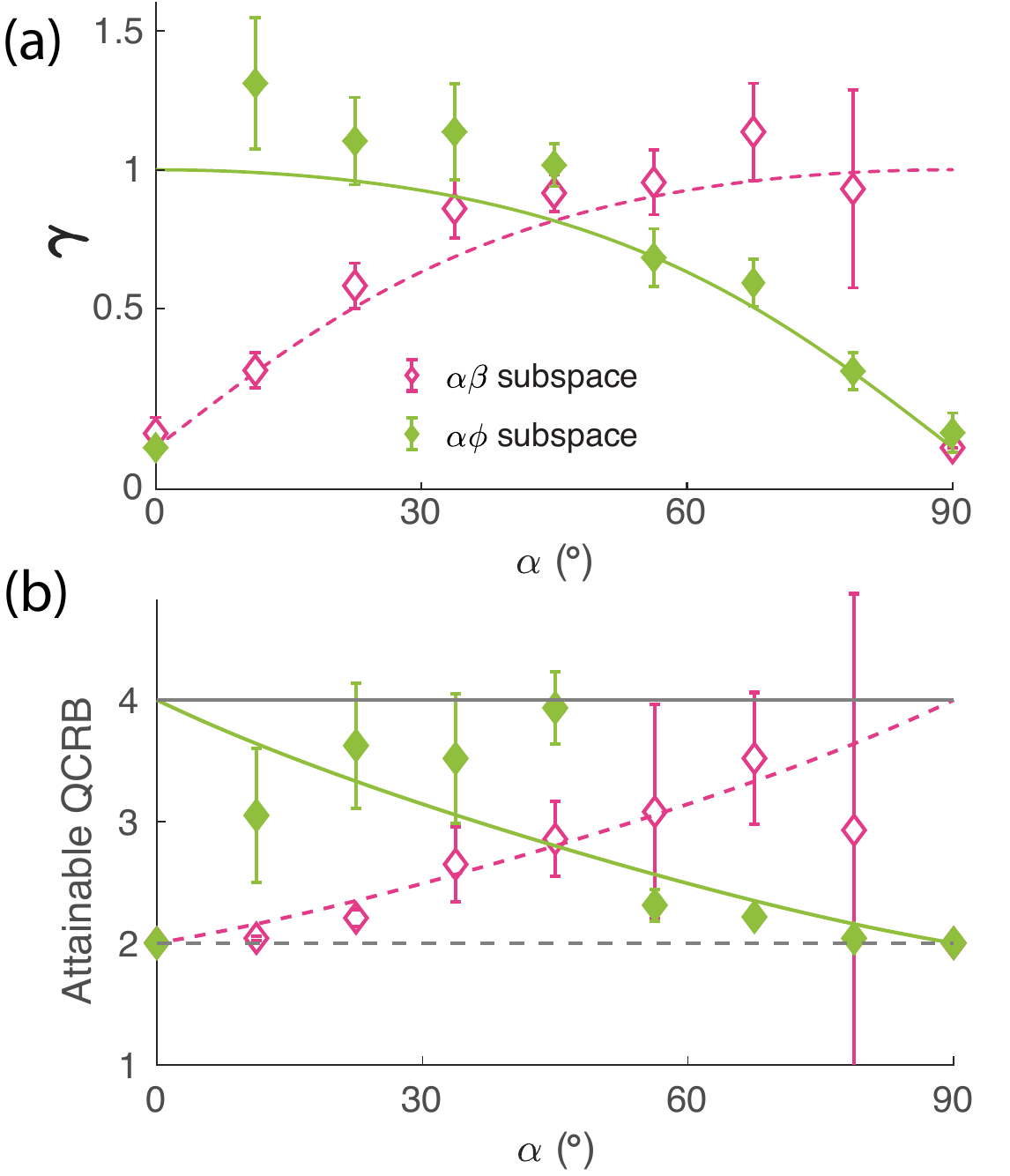}
\caption{\label{fig:fig3}.  Characterization number \textbf{(a)} and attainable QCRB  \textbf{(b)} for two-parameter estimation in the subspace of three-parameter model. The solid, colored lines are theoretical calculations in both figures. At $\alpha=0$ and at $\alpha=\pi/2$, the predicted values for $\gamma$ and attainable QCRB are 0 and 2, respectively. The solid (dashed) black lines in (b) are attainable QCRB values when $\gamma=1$ ( $\gamma=0$). }
\end{figure}

The above three-parameter model enables us to study both quasi-classical ($\gamma=0$) and coherent ($\gamma=1$) scenarios. To study cases in between, we now revisit the problem of two-parameter estimation in the subspace of the qutrit system. Specifically, we focus on the subspace spanned by parameters $(\alpha,\beta)$ and $(\alpha,\phi)$.
In Fig.~\ref{fig:fig3} we plot the characterization number $\gamma$ as well as the corresponding attainable QCRB. When $\alpha=0$ or $\frac{\pi}{2}$, the three-level system effectively reduces to a two-level system and the parameter $\phi$ or $\beta$ cannot be estimated due to the null population in $[ 0,0,1 ]^T$ or $[ 1,0,0 ]^T$, respectively. We observe  $\gamma$ varies continuously from 0 to 1 when sweeping $\alpha$ in the two subspaces. The HCRB thus ranges from 2, equal to the scalar SLD-CRB, (asymptotically) to 4, where the attainable scalar QCRB is the largest (two times larger than the SLD-CRB, as bounded by Eq.~\ref{eq:inequality}). 
These results show how $\gamma$ directly predicts the attainable QCRB for subspace models, and thus sets the estimation precision when evaluating a subset of parameters. 
The analysis above can then be applied to scenarios where there are nuisance parameters, i.e., parameters that are not of interest but still affect the precision of estimating other parameters of interest~\cite{YangCMP2019,SuzukiJPA2020,Suzuki_2020}.
We note that while we consider a three-parameter model here, the above techniques can be extended to higher-dimensional models. In Appendix~\ref{sec:Appendix4par}, we give a concrete example of a four-parameter model that can yield higher-form curvatures~\cite{Yang1978,Hasebe2014,PhysRevB.99.045154}.

The techniques described here can be directly used to evaluate the ultimate performance of practical quantum sensing applications. For example, the task of estimating parameters $\boldsymbol{\theta}$ such as those in Eq.~\ref{eq:state_3para}  can be mapped to vector magnetometry, where the estimation of $\alpha$ or $\beta, \phi$ corresponds to detecting  transverse or longitudinal magnetic fields, respectively. The unitary operator $U(\boldsymbol{\theta})$ defined by $\ket{\psi(\boldsymbol{\theta})} =U(\boldsymbol{\theta}) \ket{\psi_0}$, for example, describes the evolution of an initial resource state $\ket{\psi_0}$ under external magnetic field, the information of which is encoded in $\boldsymbol{\theta}$.

\section{Conclusion}
In conclusion, we explore the relationship between quantum geometry and quantum multi-parameter estimation. In addition to further insight into the ultimate precision limits, this relationship enables us to experimentally extract the attainable scalar quantum Cram\'er-Rao bound for a three-parameter pure state model synthesized in a single spin system. We show that the SLD-CRB is not reachable due to the non-commutativity of parameter generators that affect  the uncertainty relations of quantum mechanics. This bound is clarified through the link to quantum geometry, where the system properties are characterized not only the \textit{distance} between states (the Fubini-Study metric tensor), but also by the space curvature.
The present technique and discussions can provide tools and insights for finding optimal measurement strategies and  evaluating the ultimate precision of multi-parameter models~\cite{Albarelli2020,Yuarxiv2020,HouPRL2020,HouSA2021}.

\begin{acknowledgments}
This work is supported in part by ARO grant W911NF-11-1-0400 and NSF grant PHY1734011.
\end{acknowledgments}

\appendix

\section{Generators and quantum geometric tensor}
\label{sec:AppendixQGT}
As discussed in Sec.~\ref{sec:Theory}, the quantum geometric tensor (QGT) $\boldsymbol{\chi}$ and the quantum Fisher information (QFI) $\boldsymbol{J}$ are related. To  derive this relation, here we assume a pure state model and unitary evolutions.
Instead of starting from the parametrized state such as the three-parameter state studied in the main text, we begin with a probe state that is independent of the parameters to be estimated, and let it evolve to the target parametrized state through a unitary dynamics $U(\boldsymbol{\theta})$, i.e.,
\begin{equation}
\begin{split}
\ket{\psi(\ve{\theta})}&=\exp[-i H(\ve{\theta}) t]\ket{\psi_0}=U(\ve{\theta}) \ket{\psi_0}\\
&\rightarrow U(\ve{\theta}+d\theta_j) \ket{\psi_0}=[I+(\partial_{\theta_j}U)U^\dagger d\theta_j]U\ket{\psi_0}\\
&=\exp\{-i [i(\partial_{\theta_j}U) U^\dagger] d\theta_j\}U \ket{\psi_0}.
\end{split}
\end{equation}
Here we define the generator $\mathcal G_\mu = i(\partial_{\mu} U)U^\dagger$, which describes the derivative of the unitary operator (similar to the derivative of the state vector, but in the Heisenberg picture). In this picture, all parameters to estimate influence the probe state through the unitary evolution. 
We can then use it to calculate the geometric quantities, as shown below.

One can apply $\ket{\psi(\ve{\theta})}\rightarrow U(\ve{\theta})\ket{\psi_0}$ and plug this into the definition of the QGT:
\begin{equation}
\begin{split}
\chi_{\mu\nu}&=\bra{\partial_\mu \psi} (I-\ket{\psi}\bra{\psi})\ket{\partial_\nu\psi}\\
&\rightarrow \bra{\partial_\mu (U\psi_0)} (I-U\ket{\psi_0}\bra{\psi_0}U^\dagger)\ket{\partial_\nu (U\psi)}\\
&=\bra{\psi_0} \partial_\mu U^\dagger \partial_\nu U \ket{\psi_0}\\& - \bra{\psi_0} (\partial_\mu U^\dagger)U\ket{\psi_0} \bra{\psi_0} (U^\dagger \partial_\nu U)\ket{\psi_0}
\end{split}
\end{equation}
The real part of $\chi_{\mu\nu}$ is given by:
\begin{equation}\label{eq:metric_tensor_crude}
\begin{split}
g_{\mu\nu}=&\frac{1}{2}(\chi_{\mu\nu}+\chi_{\mu\nu}^\dagger)\\
=&\frac{1}{2}\bra{\psi_0}(\partial_\mu U^\dagger \partial_\nu U + \partial_\nu U^\dagger \partial_\mu U) \ket{\psi_0}\\
& + \bra{\psi_0} (\partial_\mu U^\dagger)U\ket{\psi_0} \bra{\psi_0} (\partial_\nu U^\dagger) U\ket{\psi_0}
\end{split}
\end{equation}
(it is easy to show the last term is real). Because $\mathcal{G}_\mu$ is Hermitian, $\mathcal{G}_\mu=\mathcal{G}_\mu^\dagger$, meaning $i(\partial_\mu U) U^\dagger = -i U (\partial_\mu U^\dagger)$. Using this equality, Eq.~\ref{eq:metric_tensor_crude} becomes
\begin{equation}\label{eq:metric_tensor_to_QFI}
\begin{split}
g_{\mu\nu}=&\frac{1}{2}\bra{\psi_0}((\partial_\mu U^\dagger) U U^\dagger \partial_\nu U + (\partial_\nu U^\dagger) U U^\dagger  (\partial_\mu U)) \ket{\psi_0}\\
& - \bra{\psi_0} (i\partial_\mu U^\dagger)U\ket{\psi_0} \bra{\psi_0} (i\partial_\nu U^\dagger) U\ket{\psi_0}\\
=&\frac{1}{2} \bra{\psi_0} \{\mathcal{G}_\mu, \mathcal{G}_\nu \}\ket{\psi_0} - \bra{\psi_0}\mathcal{G}_\mu\ket{\psi_0}\bra{\psi_0}\mathcal{G}_\nu\ket{\psi_0}\\
=&\text{cov}(\mathcal{G}_\mu,\mathcal{G}_\nu)=\frac{1}{4}J_{\mu\nu}
\end{split}
\end{equation}
Similarly, we can find the relation between Berry curvature and the generators via
\begin{equation}\label{eq:generator_Berry_curvature}
\begin{split}
F_{\mu\nu}=&i(\chi_{\mu\nu}-\chi_{\mu\nu}^\dagger)\\
=&i\bra{\psi_0}((\partial_\mu U^\dagger) U U^\dagger \partial_\nu U - (\partial_\nu U^\dagger) U U^\dagger  (\partial_\mu U)) \ket{\psi_0}\\
=&i \bra{\psi_0} [\mathcal{G}_\mu, \mathcal{G}_\nu] \ket{\psi_0}
\end{split}
\end{equation}
We now reach Eq.~(\ref{eq: relation_J_generator})-(\ref{eq: relation_F_generator}) introduced in the main text. 
\section{Multi-parameter estimation and uncertainty relations }
\label{sec:AppendixUncertainty}
In this section, we start with the well-known Heisenberg uncertainty principle and discuss its implications on the characterization number $\gamma$.
While the product uncertainty relations can be extended to general multi-operator scenarios, here we consider the three-operator case and discuss its relation with $\gamma$ in the three-parameter model introduced in the main text.
\subsection{Two parameters}
We start from the standard Heisenberg uncertainty relation:
\begin{equation}
\ave{(\Delta \hat{A})^2} \ave{(\Delta \hat{B})^2} \geq \frac{1}{4} \vert \ave{[\hat{A},\hat{B}]}\vert^2
\end{equation}
Replacing the operators $\hat A, \hat B$ with the generators of parameters $\mu,\nu$, we have
\begin{equation} 
    \ave{\Delta \mathcal G_\mu^2} \ave{\Delta \mathcal G_\nu^2} \geq  \frac{1}{4} \vert \ave{[ \mathcal G_\mu,  \mathcal G_\nu ]} \vert ^2.
\end{equation}
Using Eq.~\ref{eq:metric_tensor_to_QFI}-\ref{eq:generator_Berry_curvature} we get
\begin{equation}
   J_{\mu\mu}J_{\nu\nu} \geq 4\mathcal{F}_{\mu\nu}^2.
\end{equation}
Thus, the Berry curvature sets an upper bound to how precisely we can  estimate the parameters. It is apparent that there is a trade off between estimating $\mu$ and $\nu$ (e.g. position vs momentum).

Now we consider the Robertson-Schr\"odinger uncertainty relation, a more stringent version of the better-known Heisenberg uncertainty relation:
\begin{equation}\label{eq: Robertson-Schrodinger uncertainty relation}
\begin{split}
    \ave{\Delta \hat{A}^2}\ave{\Delta \hat{B}^2} & \geq \frac{1}{4} \vert \ave{[\hat A, \hat B ]} \vert ^2 + \frac{1}{4} \ave{  \{ \Delta \hat{A}, \Delta \hat{B} \} }^2 \\
    & = \frac{1}{4} \vert \ave{[\hat A, \hat B ]} \vert ^2 + Cov(\hat{A}, \hat{B})^2.\\
    \end{split}
\end{equation}
Replace the operators $\hat A, \hat B$ with the generators of parameters $\mu,\nu$:
\begin{equation} \label{eq: 2_generators_inequality_app}
    \ave{\Delta \mathcal G_\mu^2} \ave{\Delta \mathcal G_\nu^2} \geq  \frac{1}{4} \vert \ave{[ \mathcal G_\mu,  \mathcal G_\nu ]} \vert ^2 + Cov(\mathcal G_i\mu, \mathcal G_\nu)^2.
\end{equation}
That is, $g_{\mu\mu} g_{\nu\nu} - g_{\mu\nu}^2 \geq \frac{1}{4} F_{\mu\nu}^2$, or equivalently, as stated in the main text,
\begin{equation}\label{eq:spin1/2_F_g_relation}
    \gamma = 2\frac{\sqrt{\mathrm{det}(\boldsymbol{F})}}{\sqrt{\mathrm{det}(\boldsymbol{J})}} = \frac{1}{2}\frac{\sqrt{\mathrm{det}(\boldsymbol{F})}}{\sqrt{\mathrm{det}(\boldsymbol{g})}} \leq 1
\end{equation}
For a spin 1/2 system with state $\ket{\psi(\theta, \phi)}=\cos \frac{\theta}{2} \ket0 -\sin\frac{\theta}{2}e^{-i\phi} \ket1$, one can get equality in Eq.~\ref{eq:spin1/2_F_g_relation}, indicating that the two parameters here are informationally (maximally) exclusive.

\subsection{Three parameters}

Recall that for our spin-1 model, we have the parameterized state
\begin{equation} \label{eq:state_3para2}
    \ket{\psi(\alpha,\beta,\phi)} = \frac{1}{\sqrt{2}}\begin{pmatrix}
 \cos\alpha e^{-i\beta} \\
 -1 \\
 \sin\alpha e^{-i\phi}   \\
    \end{pmatrix},
\end{equation}
and the QGT-related matrix
\begin{equation}\label{eq:J_spin1_analytical}
\begin{split}
    & \boldsymbol{J}=4\boldsymbol{g}= 4\begin{pmatrix}
    g_{\alpha\alpha} & g_{\alpha\beta} & g_{\alpha\phi}\\
    g_{\beta\alpha} & g_{\beta\beta} & g_{\beta\phi}  \\
    g_{\phi\alpha} & g_{\phi\beta} & g_{\phi\phi}\\
    \end{pmatrix}\\
    & = 
    \begin{pmatrix}
    2 & 0 & 0\\
    0 & \cos^2 \alpha(2-\cos^2\alpha) & -\frac{1}{4}\sin^2 2\alpha  \\
    0 & -\frac{1}{4}\sin^2 2\alpha &\sin^2\alpha(2-\sin^2 \alpha)\\
    \end{pmatrix} \\
\end{split}
\end{equation}
and 
\begin{equation}\label{eq:Jtilde_spin1_analytical}
\begin{split}
    &  \boldsymbol{F}  = \frac{1}{2}
    \begin{pmatrix}
    0 & \sin (2\alpha) & -\sin (2\alpha)\\
    -\sin (2\alpha) & 0 & 0  \\
    \sin (2\alpha) & 0 & 0\\
    \end{pmatrix}. \\
\end{split}
\end{equation}
We found that now we have $\gamma= 1$ for $\alpha \in (0,\frac{\pi}{2})$, as shown in the main text.

Now we can write the quantum geometric tensor in the form of parameter generators. With Eq.~\ref{eq:metric_tensor_to_QFI} and Eq.~\ref{eq:generator_Berry_curvature}, we have
\begin{equation}\label{eq: 3_generators_equality}
\begin{split}
\ave{\Delta \mathcal G_{\alpha}^2}   & \ave{\Delta \mathcal G_{\beta}^2}\ave{\Delta \mathcal G_{\phi}^2}  = \ave{\Delta \mathcal G_{\alpha}^2} Cov(\mathcal G_\beta, \mathcal G_\phi)^2   \\
    & +\frac{1}{4} (  \ave{\Delta \mathcal G_{\phi}^2} \vert \ave{[ \mathcal G_\alpha,  \mathcal G_\beta ]} \vert ^2 	+ \ave{\Delta \mathcal G_{\beta}^2} \vert \ave{[ \mathcal G_\alpha,  \mathcal G_\phi ]} \vert ^2  \\
    &+2 Cov(\mathcal G_\beta, \mathcal G_\phi)\vert \ave{[ \mathcal G_\alpha,  \mathcal G_\beta ]} \vert\,  \vert \ave{[ \mathcal G_\alpha,  \mathcal G_\phi ]} \vert ). \\
    \end{split}
\end{equation}
We remark that while $[\mathcal G_\beta, \mathcal G_\phi] = 0$,  the parameters $\beta$ and $\phi$ cannot be simultaneously measured with low variance since
$\ave{\Delta \mathcal G_{\beta}^2}\ave{\Delta \mathcal G_{\phi}^2} \geq Cov(\mathcal G_\beta, \mathcal G_\phi)^2 $. Indeed, these two parameters are correlated with a common parameter $\alpha$ and  $[\mathcal G_\alpha, \mathcal G_{\beta (\phi)}] \neq 0$. 

In general, for three operators $\hat{A}$, $\hat{B}$ and $\hat{C}$ that satisfies $Cov(\hat{A},\hat{B})=Cov(\hat{A},\hat{C})=0$ and $[\hat B, \hat C] = 0$, the following equality can be deduced from $\gamma \leq 1$:
\begin{equation}\label{eq: 3_operator_equality}
\begin{split}
    \ave{\Delta \hat A^2}& \ave{\Delta \hat B^2}\ave{\Delta \hat C^2}  \geq \ave{\Delta \hat A^2} Cov(\hat B, \hat C)^2   \\
    & +\frac{1}{4} (  \ave{\Delta \hat C^2} \vert \ave{[ \hat A,  \hat B  ]} \vert ^2 + \ave{\Delta \hat B^2} \vert \ave{[ \hat A,  \hat C ]} \vert ^2 \\&+ 2 Cov(\hat B, \hat C)\vert \ave{[ \hat A,  \hat B ]} \vert  \vert \ave{[ \hat A,  \hat C ]} \vert ). 
    \end{split}
\end{equation}

It's possible to have the quantum Fisher information matrix in a simple diagonal form under certain parameter transformations.  The vanished off-diagonal terms then correspond to $Cov(\hat X, \hat Y) = 0$ for $ \forall \hat X, \hat Y \in \{\hat A, \hat B,\hat C \}$. Again, from the fact that $\gamma = ||i 2\boldsymbol{J}^{-1} \boldsymbol{F} ||_{\infty} \leq 1$, we reach the generalized three-operator Heisenberg uncertainty relation
\begin{align}
\label{eq: 3_operator_inequality}
    \ave{\Delta \hat A^2} &\ave{\Delta \hat B^2}\ave{\Delta \hat C^2}   
     \geq \frac{1}{4} \left(  \ave{\Delta \hat C^2} \vert \ave{[ \hat A,  \hat B  ]} \vert ^2\right.\nonumber\\
     &\left. + \ave{\Delta \hat B^2} \vert \ave{[ \hat A,  \hat C ]} \vert ^2 + \ave{\Delta \hat A^2} \vert \ave{[ \hat B,  \hat C ]} \vert ^2\right). 
\end{align}
The above product inequality has also been derived by constructing positive semi-definite matrices~\cite{SyngeRSPA1971,QinSR2016,DodonovPRA2018}. The product-form uncertainty relation for $N$ operators can be constructed similarly. To this end, we showed the connection between quantum geometry with the generalized uncertainty principles.

\section{Quantum multi-parameter Cram\'er-Rao bound}
\label{sec:AppendixMultiCramer}
Now we show the general setting of the multi-parameter estimation problem.
If the whole parameter vector is $\ve{\theta}=(\theta_0,\theta_1,\ldots,\theta_N)^T$, and the positive operator-valued measurement (POVM) set is $\{\Pi_\theta\}$, the covariance matrix of an estimator $\hat{\ve{\theta}}$ under the POVM satisfies~\cite{Liu2019}
\begin{equation}
\text{cov}(\hat{\ve{\theta}},\{\Pi_\theta\}) \geq \frac{1}{n}\mathcal{I}^{-1}(\{\Pi_\theta\}) \geq \frac{1}{n} J^{-1}=\frac{1}{4n}g^{-1},
\end{equation}
where $\mathcal{I}$ is the classical Fisher information, $J$ is the QFIM, $n$ is the repetition of the experiment, and appears naturally here as the Heisenberg scaling. The second inequality is also the (SLD) quantum  Cram\'er-Rao bound. 

The necessary and sufficient condition for the saturation of the  above SLD-CRB (the last inequality) for a pure state is to have $F_{\mu\nu}=0$, i.e., Berry curvature is a null matrix.
Since the above SLD-CRB is usually unreachable, an attainable QCRB is introduced  Ref.~\cite{Matsumoto2000,Matsumoto_2002}
\begin{align}
    C(\boldsymbol{\theta}) & = \tr{\textrm{Re}\left[\sqrt{\openone_d+ 2i \boldsymbol{J}(\boldsymbol{\theta})^{-1/2} \boldsymbol{F}(\boldsymbol{\theta}) \boldsymbol{J}(\boldsymbol{\theta})^{-1/2} }\right]^{-2}} \nonumber\\
    & = \sum_{i}\frac{2}{1+\sqrt{1-|\gamma_i|^2}},
\end{align}
where $\openone_d$ is the d-dimensional identity matrix and $\gamma_i$ is the i-th eigenvalue of $i 2\boldsymbol{J}^{-1} \boldsymbol{F}$.

The above bound coincides with the better-known Holevo quantum Cram\'er-Rao bound (HCRB) when $\gamma=$ 0 or 1. We note that HCRB could be at most twice as large as SLD-CRB~\cite{Carollo_2019}. This condition is satisfied when $\gamma=1$ in the two-parameter model above. For the three-parameter model discussed above, the ratio is $\frac{5}{3} < 2$.

\section{Relating $\gamma$ to topological structures}
\label{sec:AppendixTopo}
In this section, we discuss the relation between characterization number $\gamma$ and topological invariants that can characterize exotic monopole structures in parameter space.

In general, $\gamma$ can be expressed as the ratio between an $n-$form generalized Berry curvature $F_{\alpha\beta...}$ and the determinant of the metric tensor
\begin{equation}
    \gamma = c\frac{F_{\alpha\beta...}}{\sqrt{det(\boldsymbol{g})}}
\end{equation}
with $c$ being a constant normalization factor. 
Accordingly, $\gamma$ is U(1) gauge-invariant and for pointlike or extended monopoles in parameter space spanned by parameters $\boldsymbol{\theta} = \{ \alpha, \beta, ...\}$~\cite{GianPRL2018,chen2022}, we have $\gamma(\boldsymbol{\theta})=1,$ $\forall  \boldsymbol{\theta}$ in parameter space, indicating the largest quantumness of these monopoles. The isotropy of $\gamma$ is due to the radial nature of the field emanating from these pointlike monopole sources; when $\gamma=1$, the flux (i.e., the generalized Berry curvature) through a unit area (given by the metric tensor determinant) is unit and thus the flux integral over a closed sphere $S^n$ gives the quantized topological number, upon normalization. 

As an example, the two-parameter qubit model $\ket{\psi(\theta, \phi)}=\cos\left(\frac{\theta}{2}\right) \ket0 -\sin\left(\frac{\theta}{2}\right)e^{-i\phi} \ket1$ describes pure states on a $S^2$ sphere 
such that the integral of their Berry curvature $F_{\mu\nu}$ over the sphere yields the first Chern number, corresponding to a Dirac monopole at the origin of parameter space.

For the three-level state~\ref{eq:state_3para} studied in this work, 
it indeed corresponds to the ground state of a 4D Weyl-type Hamiltonian in spherical coordinates, that hosts a tensor monopole at the origin~\cite{chen2022}. We can rewrite the characterization number as
\begin{equation}\label{eq:relation_H_detJ}
\gamma = 2\frac{ \sqrt{J_{\phi\phi}F_{\alpha\beta}^2 +J_{\beta\beta}F_{\alpha\phi}^2 - 2 J_{\beta\phi}F_{\alpha\beta}F_{\alpha\phi}   }}{\sqrt{\det \boldsymbol{J}}} \equiv \frac{2\mathcal{H}_{\alpha\beta\phi}}{\sqrt{\det \boldsymbol{J}}} . 
\end{equation}
When $\gamma=1$, one recovers the relation between the generalized 3-from curvature $\mathcal{H}$ and metric tensor $\boldsymbol{g}$,  $\mathcal H = \frac{1}{2} \sqrt{\det \boldsymbol{J}}=4\sqrt{\det \boldsymbol{g}}$. 
The integral of the curvature $\mathcal{H}$ over the $S^3$ sphere defined by parameters $\{ \alpha, \beta, \phi \}$ yields the  Dixmier-Douady invariant, which is the topological number for a tensor monopole and its related Kalb-Ramond field~\cite{chen2022,GianPRL2018}.

As stated in the main text, the above three-parameter model enables us to study both quasi-classical ($\gamma=0$) and coherent ($\gamma=1$) cases. This model, as  well as the two-level model $\ket{\psi(\theta, \phi)}=\cos\left(\frac{\theta}{2}\right) \ket0 -\sin\left(\frac{\theta}{2}\right)e^{-i\phi} \ket1$, is one of the eigenstates of a gapless Weyl-type Hamiltonian that hosts monopoles at the origin of parameter space~\cite{Bruno2022,chen2022}, leading to  $\gamma =1$ (or 0 at special points).
To study intermediate cases where $\gamma \leq 1$, one can 
consider different geometrical models~\cite{Mera22} or, as done in this work as shown in the main text, rely on the two-parameter subspaces of the above three-parameter model where $\gamma$ is a function of $\alpha$. From the viewpoint of the Hamiltonian, this corresponds to taking a slice of the original gapless Weyl-type system, which yields a gapped topological insulator that can have  $\gamma \leq 1$.

\begin{figure}[ht]
\centering
\includegraphics[width=0.45\textwidth]{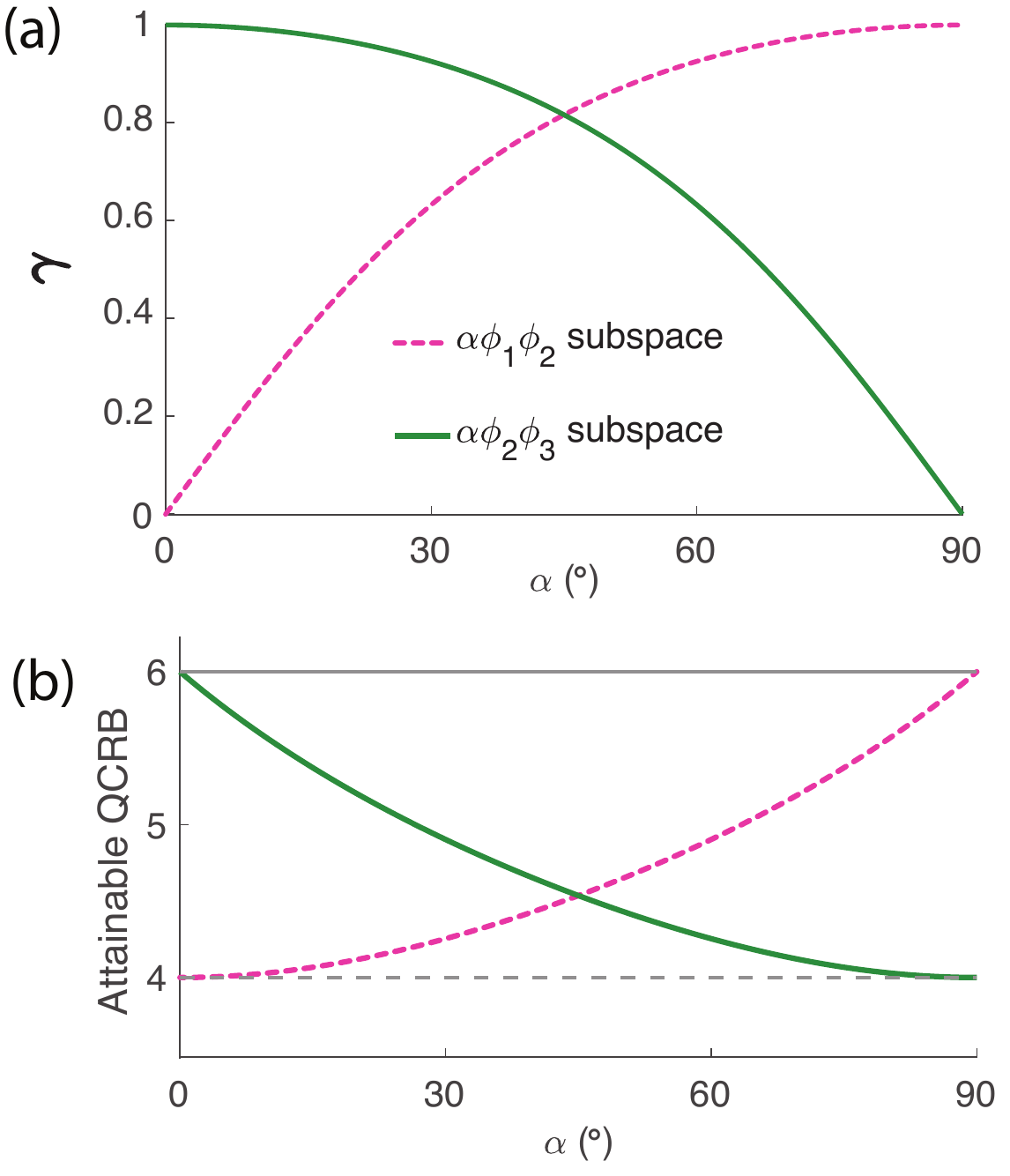}
\caption{\label{fig:spin3_2_subspace} Characterization number $\gamma$ (a) and attainable QCRB (b) values for three-parameter subspaces in the spin $\frac{3}{2}$ model Eq.~\ref{eq:spin3_2_model} for $\alpha \in (0,\pi/2)$. The solid (dashed) black lines in the right plot indicate the largest (=6) and smallest (=4) reachable attainable QCRB values for three-parameter models. }
\end{figure}

\section{Model with four parameters}
\label{sec:Appendix4par}

As an extension of the discussed three-parameter model, we  here consider the following pure state with parameters $(\alpha, \phi_1,\phi_2,\phi_3)$:
\begin{equation}\label{eq:spin3_2_model}
    \ket{\psi(\alpha,\phi_1,\phi_2,\phi_3)} = \frac{1}{\sqrt{3}}\begin{pmatrix}
 \cos\alpha e^{i\phi_1} \\
 e^{i\phi_2} \\
 \sin\alpha e^{i\phi_3}   \\
  1   \\
    \end{pmatrix},
\end{equation}
\begin{widetext}
	The quantum Fisher information and Berry curvature matrices are found to be
\begin{equation}
\begin{split}
    & \boldsymbol{J}=4\boldsymbol{g}= 4\begin{pmatrix}
    g_{\alpha\alpha} & g_{\alpha\phi_1} & g_{\alpha\phi_2} & g_{\alpha\phi_3}\\
    g_{\phi_1\alpha} & g_{\phi_1\phi_1} & g_{\phi_1\phi_2} & g_{\phi_1\phi_3}  \\
    g_{\phi_2\alpha} & g_{\phi_2\phi_1} & g_{\phi_2\phi_2} & g_{\phi_2\phi_3}\\
     g_{\phi_3\alpha} & g_{\phi_3\phi_1} & g_{\phi_3\phi_2} & g_{\phi_3\phi_3}\\
    \end{pmatrix}\\
    & = 
    \begin{pmatrix}
    4/3 & 0 & 0 & 0\\
    0 & 4/9(2\cos^2\alpha+\sin^2\alpha \cos^2 \alpha)  & -4/9\cos^2 \alpha & -4/9\sin^2\alpha \cos^2 \alpha \\
    0 & -4/9\cos^2 \alpha & 8/9 & -4/9\sin^2 \alpha \\
    0 & -4/9\sin^2\alpha \cos^2 \alpha &  -4/9\sin^2 \alpha & 4/9(2\sin^2\alpha+\sin^2\alpha \cos^2 \alpha) \\
    \end{pmatrix} \\,
\end{split}
\end{equation}
and 
\begin{equation}
\begin{split}
    & \boldsymbol{F}  = \frac{1}{2}
    \begin{pmatrix}
    0 & -4/3\cos\alpha\sin\alpha & 0 & 4/3\cos\alpha\sin\alpha\\
    4/3\cos\alpha\sin\alpha & 0 & 0  & 0\\
    0 & 0 & 0  & 0\\
    -4/3\cos\alpha\sin\alpha & 0 & 0  & 0\\
    \end{pmatrix}. \\
\end{split}
\end{equation}

Similarly, we find that $\gamma=1$ for this model when $\alpha \in (0, \frac{\pi}{2})$. The corresponding attainable QCRB for this model is $C(\boldsymbol{\theta})=6>C^{S}(\boldsymbol{\theta})=4$. We note that in this model one can find an analytical expression of $\gamma$ 
\begin{equation}
    \gamma =  2\frac{ \sqrt{F_{\alpha\phi_1}^2 det(\boldsymbol{g}(\phi_2,\phi_3)) + det(\boldsymbol{g}(\phi_1,\phi_2)) F_{\alpha\phi_3}^2 - 2 F_{\alpha\beta}F_{\alpha\phi}(J_{\phi_1 \phi_2}J_{\phi_2 \phi_3}-J_{\phi_1 \phi_3}J_{\phi_2 \phi_2})   }}{\sqrt{\det \boldsymbol{J}}}  \equiv 2 \frac{\mathcal{E}_{\alpha\phi_1\phi_2\phi_3}}{\sqrt{\det \boldsymbol{J}}}
\end{equation}
 where $\mathcal{E}_{\alpha\phi_1\phi_2\phi_3}$ is the 4-form generalized Berry curvature. Its integral over the $S^4$ sphere will yield a topological number which is directly proportional to the second Chern number of a 5D tensor monopole~\cite{Yang1978,Hasebe2014,PhysRevB.99.045154}.
\end{widetext}

Similar to the spin-1 case we discussed in the main text, we find that the $\alpha\phi_1 \phi_2$ and $\alpha\phi_2 \phi_3$ subspace can have varying $\gamma$ values thus attainable QCRB, as shown in Fig.~\ref{fig:spin3_2_subspace}.

\section{Experimental Implementation}
\label{sec:AppendixExpt}
The measurement of geometric quantities for the three-parameter model Eq.~\ref{eq:state_3para} is performed using the ground triplet state of a single nitrogen-vacancy (NV) center in diamond. While the details of the measurement process can be found in Ref.~\cite{chen2022}, here we give a brief introduction for the convenience of readers.

We used a home-built confocal microscope to initialize and measure a single NV center in an electronic grade diamond sample (Element 6, natural abundance of $^{13}$C).  A magnetic field of 490 G is applied using a permanent magnet, and the native $^{14}$N nuclear spin is thus polarized and remains in $\ket{m_I=+1}$ state during the experiments due to the excited state level anti-crossing effect. The NV of interest has a long coherence time with $T_1$ = 3.2 ms and $ T_{2,echo}>$  700 $\mu$s.

The target parameterized state Eq.~\ref{eq:state_3para} corresponds to the ground state of a 4D Weyl-type Hamiltonian, 
\begin{equation}\label{eq:Ham_parametrization}
H=H_0
\begin{pmatrix}
0 &\cos(\alpha) e^{-i\beta}&0\\
\cos(\alpha) e^{i\beta}&0&\sin(\alpha) e^{i\phi}\\
0&\sin(\alpha) e^{-i\phi}& 0
\end{pmatrix}
\end{equation}
To synthesize such a model, we apply a dual-frequency microwave to engineer the NV system Hamiltonian,
\begin{align}
    H_{NV}& = DS_z^2 + \gamma_{e} B S_z \\
    &+2\sqrt{2}\gamma_{e}[B_1 \cos(\omega_1 t + \phi_1)+B_2 \cos(\omega_2 t + \phi_2)]S_x,\nonumber
\end{align}
where $S_x, S_z$ are the spin-1 operators, $\gamma_e = 2.8$~MHz/G is the gyromagnetic ratio, and $D = 2.87$~GHz
is the zero-field energy splitting of the NV ground state. The first line above is the intrinsic NV spin Hamiltonian and the second line represents the dual-frequency microwave pulse at frequencies $\omega_1, \omega_2$ with amplitudes $B_1$, $B_2$ respectively. When both microwave frequencies are on-resonance $\omega_{1(2)} =  D \pm \gamma_{e} B$, we can reach the target Hamiltonian Eq.~\ref{eq:Ham_parametrization} with 
\begin{align}
	 \label{eq:H_DQ_frame}
    H_{NV}& = \begin{pmatrix}
    0 & B_1 e^{-i\phi_1} & 0\\
 B_1 e^{i\phi_1}& 0 & B_2 e^{i\phi_2} \\
    0 &B_2 e^{-i\phi_2} & 0 \\
    \end{pmatrix}\\\nonumber
 &   = H_0\begin{pmatrix}
   0 & \cos\alpha e^{-i\beta}&0\\
\cos\alpha e^{i\beta}&0&\sin\alpha e^{i\phi}\\
0&\sin\alpha e^{-i\phi}&0
    \end{pmatrix}
\end{align}

We then apply weak periodic modulation on the parameters $\mu, \nu\in\{\alpha,\beta,\phi\}$ to measure the quantum geometric quantities~\cite{chen2022,Ozawa2018,Yu2019}. Specifically, we consider modulations $\mu (t)=\mu_0+m_\mu \sin(\omega t + \eta)$, $\nu (t)=\nu_0+m_\nu \sin(\omega t)$, with $m_\mu, m_\nu\ll 1$ and the resulting Hamiltonian reads
\begin{equation}
     H \!\approx\!H(\alpha_0,\beta_0,\phi_0) + m_\mu \partial_\mu H \sin(\omega t + \eta) + m_\nu \partial_\nu H\sin(\omega t).
\end{equation}
To extract the metric tensor (resp.~the Berry curvature), we linearly (resp.~elliptically) modulate $\mu, \nu$ at $\eta=0$ (resp.~$\pi/2$). When the modulation frequency is tuned on resonance with $\Delta_1$ ($\Delta_2$) that is the energy gap between the state $\ket{\psi}$ in Eq.~\ref{eq:state_3para} and the other two eigenstates of Hamiltonian Eq.~\ref{eq:Ham_parametrization} $\ket{\psi_1}$ and $\ket{\psi_2}$, the parametric modulations coherently drive Rabi oscillations between $\ket{\psi}\leftrightarrow\ket{\psi_1}$ ($\ket{\psi}\leftrightarrow \ket{\psi_2}$)  and the corresponding Rabi frequencies are $\Gamma^{\mu\nu}_{1}$ ($\Gamma^{\mu\nu}_{2}$). Here the superscript $\mu\nu$ means modulations on parameter $\mu$ and $\nu$ simultaneously with amplitudes $m_\mu=m_\nu$ and we will use superscript $\mu\bar{\nu}$ to denote driving with amplitudes $m_\mu=-m_\nu$.

It can be shown that the  metric tensor and Berry curvature can be constructed from the Rabi frequencies
\begin{align}
\label{eq:g_from_Gamma}
g_{\mu\mu}&=\frac{(\Gamma_1^\mu)^2}{\Delta_1^2}+\frac{(\Gamma_2^\mu)^2}{\Delta_2^2},\\
g_{\mu\nu}&=\frac{[(\Gamma_1^{\mu\nu})^2-(\Gamma_1^{\mu\bar{\nu}})^2]}{4\Delta_1^2}+\frac{[(\Gamma_2^{\mu\nu})^2-(\Gamma_2^{\mu\bar{\nu}})^2]}{4\Delta_2^2}\quad{}\textrm{(linear)},\nonumber\\
F_{\mu\nu}&=\frac{[(\Gamma_1^{\mu\nu})^2-(\Gamma_1^{\mu\bar{\nu}})^2]}{2\Delta_1^2}+\frac{[(\Gamma_2^{\mu\nu})^2-(\Gamma_2^{\mu\bar{\nu}})^2]}{2\Delta_2^2}\quad{}\textrm{(elliptical)}.\nonumber
\end{align}
Therefore, by measuring the Rabi oscillations, we obtain the geometric quantities, as shown in Fig.2 of the main text. 

\bibliographystyle{apsrev4-1}
\bibliography{reference}

\begin{thebibliography}{43}%
\makeatletter
\providecommand \@ifxundefined [1]{%
 \@ifx{#1\undefined}
}%
\providecommand \@ifnum [1]{%
 \ifnum #1\expandafter \@firstoftwo
 \else \expandafter \@secondoftwo
 \fi
}%
\providecommand \@ifx [1]{%
 \ifx #1\expandafter \@firstoftwo
 \else \expandafter \@secondoftwo
 \fi
}%
\providecommand \natexlab [1]{#1}%
\providecommand \enquote  [1]{``#1''}%
\providecommand \bibnamefont  [1]{#1}%
\providecommand \bibfnamefont [1]{#1}%
\providecommand \citenamefont [1]{#1}%
\providecommand \href@noop [0]{\@secondoftwo}%
\providecommand \href [0]{\begingroup \@sanitize@url \@href}%
\providecommand \@href[1]{\@@startlink{#1}\@@href}%
\providecommand \@@href[1]{\endgroup#1\@@endlink}%
\providecommand \@sanitize@url [0]{\catcode `\\12\catcode `\$12\catcode
  `\&12\catcode `\#12\catcode `\^12\catcode `\_12\catcode `\%12\relax}%
\providecommand \@@startlink[1]{}%
\providecommand \@@endlink[0]{}%
\providecommand \url  [0]{\begingroup\@sanitize@url \@url }%
\providecommand \@url [1]{\endgroup\@href {#1}{\urlprefix }}%
\providecommand \urlprefix  [0]{URL }%
\providecommand \Eprint [0]{\href }%
\providecommand \doibase [0]{http://dx.doi.org/}%
\providecommand \selectlanguage [0]{\@gobble}%
\providecommand \bibinfo  [0]{\@secondoftwo}%
\providecommand \bibfield  [0]{\@secondoftwo}%
\providecommand \translation [1]{[#1]}%
\providecommand \BibitemOpen [0]{}%
\providecommand \bibitemStop [0]{}%
\providecommand \bibitemNoStop [0]{.\EOS\space}%
\providecommand \EOS [0]{\spacefactor3000\relax}%
\providecommand \BibitemShut  [1]{\csname bibitem#1\endcsname}%
\let\auto@bib@innerbib\@empty
\bibitem [{\citenamefont {Degen}\ \emph {et~al.}(2017)\citenamefont {Degen},
  \citenamefont {Reinhard},\ and\ \citenamefont
  {Cappellaro}}]{RevModPhys.89.035002}%
  \BibitemOpen
  \bibfield  {author} {\bibinfo {author} {\bibfnamefont {C.~L.}\ \bibnamefont
  {Degen}}, \bibinfo {author} {\bibfnamefont {F.}~\bibnamefont {Reinhard}}, \
  and\ \bibinfo {author} {\bibfnamefont {P.}~\bibnamefont {Cappellaro}},\
  }\href {\doibase 10.1103/RevModPhys.89.035002} {\bibfield  {journal}
  {\bibinfo  {journal} {Rev. Mod. Phys.}\ }\textbf {\bibinfo {volume} {89}},\
  \bibinfo {pages} {035002} (\bibinfo {year} {2017})}\BibitemShut {NoStop}%
\bibitem [{\citenamefont {Marchiori}\ \emph {et~al.}(2022)\citenamefont
  {Marchiori}, \citenamefont {Ceccarelli}, \citenamefont {Rossi}, \citenamefont
  {Lorenzelli}, \citenamefont {Degen},\ and\ \citenamefont
  {Poggio}}]{Estefani2022}%
  \BibitemOpen
  \bibfield  {author} {\bibinfo {author} {\bibfnamefont {E.}~\bibnamefont
  {Marchiori}}, \bibinfo {author} {\bibfnamefont {L.}~\bibnamefont
  {Ceccarelli}}, \bibinfo {author} {\bibfnamefont {N.}~\bibnamefont {Rossi}},
  \bibinfo {author} {\bibfnamefont {L.}~\bibnamefont {Lorenzelli}}, \bibinfo
  {author} {\bibfnamefont {C.~L.}\ \bibnamefont {Degen}}, \ and\ \bibinfo
  {author} {\bibfnamefont {M.}~\bibnamefont {Poggio}},\ }\href {\doibase
  10.1038/s42254-021-00380-9} {\bibfield  {journal} {\bibinfo  {journal}
  {Nature Reviews Physics}\ }\textbf {\bibinfo {volume} {4}},\ \bibinfo {pages}
  {49} (\bibinfo {year} {2022})}\BibitemShut {NoStop}%
\bibitem [{\citenamefont {Thiel}\ \emph {et~al.}(2019)\citenamefont {Thiel},
  \citenamefont {Wang}, \citenamefont {Tschudin}, \citenamefont {Rohner},
  \citenamefont {Guti{\'e}rrez-Lezama}, \citenamefont {Ubrig}, \citenamefont
  {Gibertini}, \citenamefont {Giannini}, \citenamefont {Morpurgo},\ and\
  \citenamefont {Maletinsky}}]{Thiel2019}%
  \BibitemOpen
  \bibfield  {author} {\bibinfo {author} {\bibfnamefont {L.}~\bibnamefont
  {Thiel}}, \bibinfo {author} {\bibfnamefont {Z.}~\bibnamefont {Wang}},
  \bibinfo {author} {\bibfnamefont {M.~A.}\ \bibnamefont {Tschudin}}, \bibinfo
  {author} {\bibfnamefont {D.}~\bibnamefont {Rohner}}, \bibinfo {author}
  {\bibfnamefont {I.}~\bibnamefont {Guti{\'e}rrez-Lezama}}, \bibinfo {author}
  {\bibfnamefont {N.}~\bibnamefont {Ubrig}}, \bibinfo {author} {\bibfnamefont
  {M.}~\bibnamefont {Gibertini}}, \bibinfo {author} {\bibfnamefont
  {E.}~\bibnamefont {Giannini}}, \bibinfo {author} {\bibfnamefont {A.~F.}\
  \bibnamefont {Morpurgo}}, \ and\ \bibinfo {author} {\bibfnamefont
  {P.}~\bibnamefont {Maletinsky}},\ }\href {\doibase 10.1126/science.aav6926}
  {\bibfield  {journal} {\bibinfo  {journal} {Science}\ }\textbf {\bibinfo
  {volume} {364}},\ \bibinfo {pages} {973} (\bibinfo {year}
  {2019})}\BibitemShut {NoStop}%
\bibitem [{\citenamefont {Yip}\ \emph {et~al.}(2019)\citenamefont {Yip},
  \citenamefont {Ho}, \citenamefont {Yu}, \citenamefont {Chen}, \citenamefont
  {Zhang}, \citenamefont {Kasahara}, \citenamefont {Mizukami}, \citenamefont
  {Shibauchi}, \citenamefont {Matsuda}, \citenamefont {Goh},\ and\
  \citenamefont {Yang}}]{Yip2019}%
  \BibitemOpen
  \bibfield  {author} {\bibinfo {author} {\bibfnamefont {K.~Y.}\ \bibnamefont
  {Yip}}, \bibinfo {author} {\bibfnamefont {K.~O.}\ \bibnamefont {Ho}},
  \bibinfo {author} {\bibfnamefont {K.~Y.}\ \bibnamefont {Yu}}, \bibinfo
  {author} {\bibfnamefont {Y.}~\bibnamefont {Chen}}, \bibinfo {author}
  {\bibfnamefont {W.}~\bibnamefont {Zhang}}, \bibinfo {author} {\bibfnamefont
  {S.}~\bibnamefont {Kasahara}}, \bibinfo {author} {\bibfnamefont
  {Y.}~\bibnamefont {Mizukami}}, \bibinfo {author} {\bibfnamefont
  {T.}~\bibnamefont {Shibauchi}}, \bibinfo {author} {\bibfnamefont
  {Y.}~\bibnamefont {Matsuda}}, \bibinfo {author} {\bibfnamefont {S.~K.}\
  \bibnamefont {Goh}}, \ and\ \bibinfo {author} {\bibfnamefont
  {S.}~\bibnamefont {Yang}},\ }\href {\doibase 10.1126/science.aaw4278}
  {\bibfield  {journal} {\bibinfo  {journal} {Science}\ }\textbf {\bibinfo
  {volume} {366}},\ \bibinfo {pages} {1355} (\bibinfo {year}
  {2019})}\BibitemShut {NoStop}%
\bibitem [{\citenamefont {Lesik}\ \emph {et~al.}(2019)\citenamefont {Lesik},
  \citenamefont {Plisson}, \citenamefont {Toraille}, \citenamefont {Renaud},
  \citenamefont {Occelli}, \citenamefont {Schmidt}, \citenamefont {Salord},
  \citenamefont {Delobbe}, \citenamefont {Debuisschert}, \citenamefont
  {Rondin}, \citenamefont {Loubeyre},\ and\ \citenamefont {Roch}}]{Lesik2019}%
  \BibitemOpen
  \bibfield  {author} {\bibinfo {author} {\bibfnamefont {M.}~\bibnamefont
  {Lesik}}, \bibinfo {author} {\bibfnamefont {T.}~\bibnamefont {Plisson}},
  \bibinfo {author} {\bibfnamefont {L.}~\bibnamefont {Toraille}}, \bibinfo
  {author} {\bibfnamefont {J.}~\bibnamefont {Renaud}}, \bibinfo {author}
  {\bibfnamefont {F.}~\bibnamefont {Occelli}}, \bibinfo {author} {\bibfnamefont
  {M.}~\bibnamefont {Schmidt}}, \bibinfo {author} {\bibfnamefont
  {O.}~\bibnamefont {Salord}}, \bibinfo {author} {\bibfnamefont
  {A.}~\bibnamefont {Delobbe}}, \bibinfo {author} {\bibfnamefont
  {T.}~\bibnamefont {Debuisschert}}, \bibinfo {author} {\bibfnamefont
  {L.}~\bibnamefont {Rondin}}, \bibinfo {author} {\bibfnamefont
  {P.}~\bibnamefont {Loubeyre}}, \ and\ \bibinfo {author} {\bibfnamefont
  {J.-F.}\ \bibnamefont {Roch}},\ }\href {\doibase 10.1126/science.aaw4329}
  {\bibfield  {journal} {\bibinfo  {journal} {Science}\ }\textbf {\bibinfo
  {volume} {366}},\ \bibinfo {pages} {1359} (\bibinfo {year}
  {2019})}\BibitemShut {NoStop}%
\bibitem [{\citenamefont {Hsieh}\ \emph {et~al.}(2019)\citenamefont {Hsieh},
  \citenamefont {Bhattacharyya}, \citenamefont {Zu}, \citenamefont {Mittiga},
  \citenamefont {Smart}, \citenamefont {Machado}, \citenamefont {Kobrin},
  \citenamefont {Hohn}, \citenamefont {Rui}, \citenamefont {Kamrani},
  \citenamefont {Chatterjee}, \citenamefont {Choi}, \citenamefont {Zaletel},
  \citenamefont {Struzhkin}, \citenamefont {Moore}, \citenamefont {Levitas},
  \citenamefont {Jeanloz},\ and\ \citenamefont {Yao}}]{Hsieh2019}%
  \BibitemOpen
  \bibfield  {author} {\bibinfo {author} {\bibfnamefont {S.}~\bibnamefont
  {Hsieh}}, \bibinfo {author} {\bibfnamefont {P.}~\bibnamefont
  {Bhattacharyya}}, \bibinfo {author} {\bibfnamefont {C.}~\bibnamefont {Zu}},
  \bibinfo {author} {\bibfnamefont {T.}~\bibnamefont {Mittiga}}, \bibinfo
  {author} {\bibfnamefont {T.~J.}\ \bibnamefont {Smart}}, \bibinfo {author}
  {\bibfnamefont {F.}~\bibnamefont {Machado}}, \bibinfo {author} {\bibfnamefont
  {B.}~\bibnamefont {Kobrin}}, \bibinfo {author} {\bibfnamefont {T.~O.}\
  \bibnamefont {Hohn}}, \bibinfo {author} {\bibfnamefont {N.~Z.}\ \bibnamefont
  {Rui}}, \bibinfo {author} {\bibfnamefont {M.}~\bibnamefont {Kamrani}},
  \bibinfo {author} {\bibfnamefont {S.}~\bibnamefont {Chatterjee}}, \bibinfo
  {author} {\bibfnamefont {S.}~\bibnamefont {Choi}}, \bibinfo {author}
  {\bibfnamefont {M.}~\bibnamefont {Zaletel}}, \bibinfo {author} {\bibfnamefont
  {V.~V.}\ \bibnamefont {Struzhkin}}, \bibinfo {author} {\bibfnamefont {J.~E.}\
  \bibnamefont {Moore}}, \bibinfo {author} {\bibfnamefont {V.~I.}\ \bibnamefont
  {Levitas}}, \bibinfo {author} {\bibfnamefont {R.}~\bibnamefont {Jeanloz}}, \
  and\ \bibinfo {author} {\bibfnamefont {N.~Y.}\ \bibnamefont {Yao}},\ }\href
  {\doibase 10.1126/science.aaw4352} {\bibfield  {journal} {\bibinfo  {journal}
  {Science}\ }\textbf {\bibinfo {volume} {366}},\ \bibinfo {pages} {1349}
  (\bibinfo {year} {2019})}\BibitemShut {NoStop}%
\bibitem [{\citenamefont {Kucsko}\ \emph {et~al.}(2013)\citenamefont {Kucsko},
  \citenamefont {Maurer}, \citenamefont {Yao}, \citenamefont {Kubo},
  \citenamefont {Noh}, \citenamefont {Lo}, \citenamefont {Park},\ and\
  \citenamefont {Lukin}}]{Kucsko2013}%
  \BibitemOpen
  \bibfield  {author} {\bibinfo {author} {\bibfnamefont {G.}~\bibnamefont
  {Kucsko}}, \bibinfo {author} {\bibfnamefont {P.~C.}\ \bibnamefont {Maurer}},
  \bibinfo {author} {\bibfnamefont {N.~Y.}\ \bibnamefont {Yao}}, \bibinfo
  {author} {\bibfnamefont {M.}~\bibnamefont {Kubo}}, \bibinfo {author}
  {\bibfnamefont {H.~J.}\ \bibnamefont {Noh}}, \bibinfo {author} {\bibfnamefont
  {P.~K.}\ \bibnamefont {Lo}}, \bibinfo {author} {\bibfnamefont
  {H.}~\bibnamefont {Park}}, \ and\ \bibinfo {author} {\bibfnamefont {M.~D.}\
  \bibnamefont {Lukin}},\ }\href {\doibase 10.1038/nature12373} {\bibfield
  {journal} {\bibinfo  {journal} {Nature}\ }\textbf {\bibinfo {volume} {500}},\
  \bibinfo {pages} {54} (\bibinfo {year} {2013})}\BibitemShut {NoStop}%
\bibitem [{\citenamefont {Choi}\ \emph {et~al.}(2020)\citenamefont {Choi},
  \citenamefont {Zhou}, \citenamefont {Landig}, \citenamefont {Wu},
  \citenamefont {Yu}, \citenamefont {Stetina}, \citenamefont {Kucsko},
  \citenamefont {Mango}, \citenamefont {Needleman}, \citenamefont {Samuel},
  \citenamefont {Maurer}, \citenamefont {Park},\ and\ \citenamefont
  {Lukin}}]{Choi2020}%
  \BibitemOpen
  \bibfield  {author} {\bibinfo {author} {\bibfnamefont {J.}~\bibnamefont
  {Choi}}, \bibinfo {author} {\bibfnamefont {H.}~\bibnamefont {Zhou}}, \bibinfo
  {author} {\bibfnamefont {R.}~\bibnamefont {Landig}}, \bibinfo {author}
  {\bibfnamefont {H.-Y.}\ \bibnamefont {Wu}}, \bibinfo {author} {\bibfnamefont
  {X.}~\bibnamefont {Yu}}, \bibinfo {author} {\bibfnamefont {S.~E.~V.}\
  \bibnamefont {Stetina}}, \bibinfo {author} {\bibfnamefont {G.}~\bibnamefont
  {Kucsko}}, \bibinfo {author} {\bibfnamefont {S.~E.}\ \bibnamefont {Mango}},
  \bibinfo {author} {\bibfnamefont {D.~J.}\ \bibnamefont {Needleman}}, \bibinfo
  {author} {\bibfnamefont {A.~D.~T.}\ \bibnamefont {Samuel}}, \bibinfo {author}
  {\bibfnamefont {P.~C.}\ \bibnamefont {Maurer}}, \bibinfo {author}
  {\bibfnamefont {H.}~\bibnamefont {Park}}, \ and\ \bibinfo {author}
  {\bibfnamefont {M.~D.}\ \bibnamefont {Lukin}},\ }\href {\doibase
  10.1073/pnas.1922730117} {\bibfield  {journal} {\bibinfo  {journal}
  {Proceedings of the National Academy of Sciences}\ }\textbf {\bibinfo
  {volume} {117}},\ \bibinfo {pages} {14636} (\bibinfo {year}
  {2020})}\BibitemShut {NoStop}%
\bibitem [{\citenamefont {Li}\ \emph {et~al.}(2022)\citenamefont {Li},
  \citenamefont {Soleyman}, \citenamefont {Kohandel},\ and\ \citenamefont
  {Cappellaro}}]{Li2022}%
  \BibitemOpen
  \bibfield  {author} {\bibinfo {author} {\bibfnamefont {C.}~\bibnamefont
  {Li}}, \bibinfo {author} {\bibfnamefont {R.}~\bibnamefont {Soleyman}},
  \bibinfo {author} {\bibfnamefont {M.}~\bibnamefont {Kohandel}}, \ and\
  \bibinfo {author} {\bibfnamefont {P.}~\bibnamefont {Cappellaro}},\ }\bibfield
   {booktitle} {\emph {\bibinfo {booktitle} {Nano Letters}},\ }\href {\doibase
  10.1021/acs.nanolett.1c02868} {\bibfield  {journal} {\bibinfo  {journal}
  {Nano Letters}\ }\textbf {\bibinfo {volume} {22}},\ \bibinfo {pages} {43}
  (\bibinfo {year} {2022})}\BibitemShut {NoStop}%
\bibitem [{\citenamefont {Helstrom}(1967)}]{Helstrom1967}%
  \BibitemOpen
  \bibfield  {author} {\bibinfo {author} {\bibfnamefont {C.}~\bibnamefont
  {Helstrom}},\ }\href {\doibase https://doi.org/10.1016/0375-9601(67)90366-0}
  {\bibfield  {journal} {\bibinfo  {journal} {Physics Letters A}\ }\textbf
  {\bibinfo {volume} {25}},\ \bibinfo {pages} {101} (\bibinfo {year}
  {1967})}\BibitemShut {NoStop}%
\bibitem [{\citenamefont {Yuen}\ and\ \citenamefont {Lax}(1973)}]{Yuen1973}%
  \BibitemOpen
  \bibfield  {author} {\bibinfo {author} {\bibfnamefont {H.}~\bibnamefont
  {Yuen}}\ and\ \bibinfo {author} {\bibfnamefont {M.}~\bibnamefont {Lax}},\
  }\href {\doibase 10.1109/TIT.1973.1055103} {\bibfield  {journal} {\bibinfo
  {journal} {IEEE Transactions on Information Theory}\ }\textbf {\bibinfo
  {volume} {19}},\ \bibinfo {pages} {740} (\bibinfo {year} {1973})}\BibitemShut
  {NoStop}%
\bibitem [{\citenamefont {Belavkin}(1976)}]{Belavkin1976}%
  \BibitemOpen
  \bibfield  {author} {\bibinfo {author} {\bibfnamefont {V.~P.}\ \bibnamefont
  {Belavkin}},\ }\href {\doibase 10.1007/BF01032091} {\bibfield  {journal}
  {\bibinfo  {journal} {Theoretical and Mathematical Physics}\ }\textbf
  {\bibinfo {volume} {26}},\ \bibinfo {pages} {213} (\bibinfo {year}
  {1976})}\BibitemShut {NoStop}%
\bibitem [{\citenamefont {Holevo}(1977)}]{Holevo1977}%
  \BibitemOpen
  \bibfield  {author} {\bibinfo {author} {\bibfnamefont {A.}~\bibnamefont
  {Holevo}},\ }\href {\doibase https://doi.org/10.1016/0034-4877(77)90009-X}
  {\bibfield  {journal} {\bibinfo  {journal} {Reports on Mathematical Physics}\
  }\textbf {\bibinfo {volume} {12}},\ \bibinfo {pages} {251} (\bibinfo {year}
  {1977})}\BibitemShut {NoStop}%
\bibitem [{\citenamefont {Paris}(2009)}]{PARIS2009}%
  \BibitemOpen
  \bibfield  {author} {\bibinfo {author} {\bibfnamefont {M.~G.~A.}\
  \bibnamefont {Paris}},\ }\href {\doibase 10.1142/s0219749909004839}
  {\bibfield  {journal} {\bibinfo  {journal} {International Journal of Quantum
  Information}\ }\textbf {\bibinfo {volume} {07}},\ \bibinfo {pages} {125}
  (\bibinfo {year} {2009})}\BibitemShut {NoStop}%
\bibitem [{\citenamefont {Carollo}\ \emph {et~al.}(2019)\citenamefont
  {Carollo}, \citenamefont {Spagnolo}, \citenamefont {Dubkov},\ and\
  \citenamefont {Valenti}}]{Carollo_2019}%
  \BibitemOpen
  \bibfield  {author} {\bibinfo {author} {\bibfnamefont {A.}~\bibnamefont
  {Carollo}}, \bibinfo {author} {\bibfnamefont {B.}~\bibnamefont {Spagnolo}},
  \bibinfo {author} {\bibfnamefont {A.~A.}\ \bibnamefont {Dubkov}}, \ and\
  \bibinfo {author} {\bibfnamefont {D.}~\bibnamefont {Valenti}},\ }\href
  {\doibase 10.1088/1742-5468/ab3ccb} {\bibfield  {journal} {\bibinfo
  {journal} {Journal of Statistical Mechanics: Theory and Experiment}\ }\textbf
  {\bibinfo {volume} {2019}},\ \bibinfo {pages} {094010} (\bibinfo {year}
  {2019})}\BibitemShut {NoStop}%
\bibitem [{\citenamefont {Liu}\ \emph {et~al.}(2019)\citenamefont {Liu},
  \citenamefont {Yuan}, \citenamefont {Lu},\ and\ \citenamefont
  {Wang}}]{Liu2019}%
  \BibitemOpen
  \bibfield  {author} {\bibinfo {author} {\bibfnamefont {J.}~\bibnamefont
  {Liu}}, \bibinfo {author} {\bibfnamefont {H.}~\bibnamefont {Yuan}}, \bibinfo
  {author} {\bibfnamefont {X.-M.}\ \bibnamefont {Lu}}, \ and\ \bibinfo {author}
  {\bibfnamefont {X.}~\bibnamefont {Wang}},\ }\href {\doibase
  10.1088/1751-8121/ab5d4d} {\bibfield  {journal} {\bibinfo  {journal} {Journal
  of Physics A: Mathematical and Theoretical}\ }\textbf {\bibinfo {volume}
  {53}},\ \bibinfo {pages} {023001} (\bibinfo {year} {2019})}\BibitemShut
  {NoStop}%
\bibitem [{\citenamefont {Albarelli}\ \emph {et~al.}(2020)\citenamefont
  {Albarelli}, \citenamefont {Barbieri}, \citenamefont {Genoni},\ and\
  \citenamefont {Gianani}}]{Albarelli2020}%
  \BibitemOpen
  \bibfield  {author} {\bibinfo {author} {\bibfnamefont {F.}~\bibnamefont
  {Albarelli}}, \bibinfo {author} {\bibfnamefont {M.}~\bibnamefont {Barbieri}},
  \bibinfo {author} {\bibfnamefont {M.}~\bibnamefont {Genoni}}, \ and\ \bibinfo
  {author} {\bibfnamefont {I.}~\bibnamefont {Gianani}},\ }\href {\doibase
  https://doi.org/10.1016/j.physleta.2020.126311} {\bibfield  {journal}
  {\bibinfo  {journal} {Physics Letters A}\ }\textbf {\bibinfo {volume}
  {384}},\ \bibinfo {pages} {126311} (\bibinfo {year} {2020})}\BibitemShut
  {NoStop}%
\bibitem [{\citenamefont {Braunstein}\ and\ \citenamefont
  {Caves}(1994)}]{BraunsteinPRL1994}%
  \BibitemOpen
  \bibfield  {author} {\bibinfo {author} {\bibfnamefont {S.~L.}\ \bibnamefont
  {Braunstein}}\ and\ \bibinfo {author} {\bibfnamefont {C.~M.}\ \bibnamefont
  {Caves}},\ }\href {\doibase 10.1103/PhysRevLett.72.3439} {\bibfield
  {journal} {\bibinfo  {journal} {Phys. Rev. Lett.}\ }\textbf {\bibinfo
  {volume} {72}},\ \bibinfo {pages} {3439} (\bibinfo {year}
  {1994})}\BibitemShut {NoStop}%
\bibitem [{\citenamefont {Guo}\ \emph {et~al.}(2016)\citenamefont {Guo},
  \citenamefont {Zhong}, \citenamefont {Jing}, \citenamefont {Fu},\ and\
  \citenamefont {Wang}}]{GuoPRA2016}%
  \BibitemOpen
  \bibfield  {author} {\bibinfo {author} {\bibfnamefont {W.}~\bibnamefont
  {Guo}}, \bibinfo {author} {\bibfnamefont {W.}~\bibnamefont {Zhong}}, \bibinfo
  {author} {\bibfnamefont {X.-X.}\ \bibnamefont {Jing}}, \bibinfo {author}
  {\bibfnamefont {L.-B.}\ \bibnamefont {Fu}}, \ and\ \bibinfo {author}
  {\bibfnamefont {X.}~\bibnamefont {Wang}},\ }\href {\doibase
  10.1103/PhysRevA.93.042115} {\bibfield  {journal} {\bibinfo  {journal} {Phys.
  Rev. A}\ }\textbf {\bibinfo {volume} {93}},\ \bibinfo {pages} {042115}
  (\bibinfo {year} {2016})}\BibitemShut {NoStop}%
\bibitem [{\citenamefont {Kolodrubetz}\ \emph {et~al.}(2017)\citenamefont
  {Kolodrubetz}, \citenamefont {Sels}, \citenamefont {Mehta},\ and\
  \citenamefont {Polkovnikov}}]{KolodrubetzPR2017}%
  \BibitemOpen
  \bibfield  {author} {\bibinfo {author} {\bibfnamefont {M.}~\bibnamefont
  {Kolodrubetz}}, \bibinfo {author} {\bibfnamefont {D.}~\bibnamefont {Sels}},
  \bibinfo {author} {\bibfnamefont {P.}~\bibnamefont {Mehta}}, \ and\ \bibinfo
  {author} {\bibfnamefont {A.}~\bibnamefont {Polkovnikov}},\ }\href {\doibase
  https://doi.org/10.1016/j.physrep.2017.07.001} {\bibfield  {journal}
  {\bibinfo  {journal} {Physics Reports}\ }\textbf {\bibinfo {volume} {697}},\
  \bibinfo {pages} {1} (\bibinfo {year} {2017})},\ \bibinfo {note} {geometry
  and non-adiabatic response in quantum and classical systems}\BibitemShut
  {NoStop}%
\bibitem [{\citenamefont {Matsumoto}(2000)}]{Matsumoto2000}%
  \BibitemOpen
  \bibfield  {author} {\bibinfo {author} {\bibfnamefont {K.}~\bibnamefont
  {Matsumoto}},\ }\href {https://arxiv.org/abs/quant-ph/0006076} {\bibfield
  {journal} {\bibinfo  {journal} {arXiv:quant-ph/0006076}\ } (\bibinfo {year}
  {2000})}\BibitemShut {NoStop}%
\bibitem [{\citenamefont {Matsumoto}(2002)}]{Matsumoto_2002}%
  \BibitemOpen
  \bibfield  {author} {\bibinfo {author} {\bibfnamefont {K.}~\bibnamefont
  {Matsumoto}},\ }\href {\doibase 10.1088/0305-4470/35/13/307} {\bibfield
  {journal} {\bibinfo  {journal} {Journal of Physics A: Mathematical and
  General}\ }\textbf {\bibinfo {volume} {35}},\ \bibinfo {pages} {3111}
  (\bibinfo {year} {2002})}\BibitemShut {NoStop}%
\bibitem [{\citenamefont {Robertson}(1934)}]{Robertson1934}%
  \BibitemOpen
  \bibfield  {author} {\bibinfo {author} {\bibfnamefont {H.~P.}\ \bibnamefont
  {Robertson}},\ }\href {\doibase 10.1103/PhysRev.46.794} {\bibfield  {journal}
  {\bibinfo  {journal} {Phys. Rev.}\ }\textbf {\bibinfo {volume} {46}},\
  \bibinfo {pages} {794} (\bibinfo {year} {1934})}\BibitemShut {NoStop}%
\bibitem [{\citenamefont {Ozawa}\ and\ \citenamefont
  {Mera}(2021)}]{Ozawa_BrunoPRB2021}%
  \BibitemOpen
  \bibfield  {author} {\bibinfo {author} {\bibfnamefont {T.}~\bibnamefont
  {Ozawa}}\ and\ \bibinfo {author} {\bibfnamefont {B.}~\bibnamefont {Mera}},\
  }\href {\doibase 10.1103/PhysRevB.104.045103} {\bibfield  {journal} {\bibinfo
   {journal} {Phys. Rev. B}\ }\textbf {\bibinfo {volume} {104}},\ \bibinfo
  {pages} {045103} (\bibinfo {year} {2021})}\BibitemShut {NoStop}%
\bibitem [{\citenamefont {Mera}\ and\ \citenamefont
  {Ozawa}(2021)}]{Bruno_OzawaPRB2021}%
  \BibitemOpen
  \bibfield  {author} {\bibinfo {author} {\bibfnamefont {B.}~\bibnamefont
  {Mera}}\ and\ \bibinfo {author} {\bibfnamefont {T.}~\bibnamefont {Ozawa}},\
  }\href {\doibase 10.1103/PhysRevB.104.045104} {\bibfield  {journal} {\bibinfo
   {journal} {Phys. Rev. B}\ }\textbf {\bibinfo {volume} {104}},\ \bibinfo
  {pages} {045104} (\bibinfo {year} {2021})}\BibitemShut {NoStop}%
\bibitem [{\citenamefont {Mera}\ \emph
  {et~al.}(2022{\natexlab{a}})\citenamefont {Mera}, \citenamefont {Zhang},\
  and\ \citenamefont {Goldman}}]{Bruno2022}%
  \BibitemOpen
  \bibfield  {author} {\bibinfo {author} {\bibfnamefont {B.}~\bibnamefont
  {Mera}}, \bibinfo {author} {\bibfnamefont {A.}~\bibnamefont {Zhang}}, \ and\
  \bibinfo {author} {\bibfnamefont {N.}~\bibnamefont {Goldman}},\ }\href
  {\doibase 10.21468/SciPostPhys.12.1.018} {\bibfield  {journal} {\bibinfo
  {journal} {SciPost Phys.}\ }\textbf {\bibinfo {volume} {12}},\ \bibinfo
  {pages} {18} (\bibinfo {year} {2022}{\natexlab{a}})}\BibitemShut {NoStop}%
\bibitem [{\citenamefont {Chen}\ \emph {et~al.}(2022)\citenamefont {Chen},
  \citenamefont {Li}, \citenamefont {Palumbo}, \citenamefont {Zhu},
  \citenamefont {Goldman},\ and\ \citenamefont {Cappellaro}}]{chen2022}%
  \BibitemOpen
  \bibfield  {author} {\bibinfo {author} {\bibfnamefont {M.}~\bibnamefont
  {Chen}}, \bibinfo {author} {\bibfnamefont {C.}~\bibnamefont {Li}}, \bibinfo
  {author} {\bibfnamefont {G.}~\bibnamefont {Palumbo}}, \bibinfo {author}
  {\bibfnamefont {Y.-Q.}\ \bibnamefont {Zhu}}, \bibinfo {author} {\bibfnamefont
  {N.}~\bibnamefont {Goldman}}, \ and\ \bibinfo {author} {\bibfnamefont
  {P.}~\bibnamefont {Cappellaro}},\ }\href {\doibase 10.1126/science.abe6437}
  {\bibfield  {journal} {\bibinfo  {journal} {Science}\ }\textbf {\bibinfo
  {volume} {375}},\ \bibinfo {pages} {1017} (\bibinfo {year}
  {2022})}\BibitemShut {NoStop}%
\bibitem [{\citenamefont {Ozawa}\ and\ \citenamefont
  {Goldman}(2018)}]{Ozawa2018}%
  \BibitemOpen
  \bibfield  {author} {\bibinfo {author} {\bibfnamefont {T.}~\bibnamefont
  {Ozawa}}\ and\ \bibinfo {author} {\bibfnamefont {N.}~\bibnamefont
  {Goldman}},\ }\href {\doibase 10.1103/PhysRevB.97.201117} {\bibfield
  {journal} {\bibinfo  {journal} {Phys. Rev. B}\ }\textbf {\bibinfo {volume}
  {97}},\ \bibinfo {pages} {201117} (\bibinfo {year} {2018})}\BibitemShut
  {NoStop}%
\bibitem [{\citenamefont {Yu}\ \emph {et~al.}(2019)\citenamefont {Yu},
  \citenamefont {Yang}, \citenamefont {Gong}, \citenamefont {Cao},
  \citenamefont {Lu}, \citenamefont {Liu}, \citenamefont {Zhang}, \citenamefont
  {Plenio}, \citenamefont {Jelezko}, \citenamefont {Ozawa}, \citenamefont
  {Goldman},\ and\ \citenamefont {Cai}}]{Yu2019}%
  \BibitemOpen
  \bibfield  {author} {\bibinfo {author} {\bibfnamefont {M.}~\bibnamefont
  {Yu}}, \bibinfo {author} {\bibfnamefont {P.}~\bibnamefont {Yang}}, \bibinfo
  {author} {\bibfnamefont {M.}~\bibnamefont {Gong}}, \bibinfo {author}
  {\bibfnamefont {Q.}~\bibnamefont {Cao}}, \bibinfo {author} {\bibfnamefont
  {Q.}~\bibnamefont {Lu}}, \bibinfo {author} {\bibfnamefont {H.}~\bibnamefont
  {Liu}}, \bibinfo {author} {\bibfnamefont {S.}~\bibnamefont {Zhang}}, \bibinfo
  {author} {\bibfnamefont {M.~B.}\ \bibnamefont {Plenio}}, \bibinfo {author}
  {\bibfnamefont {F.}~\bibnamefont {Jelezko}}, \bibinfo {author} {\bibfnamefont
  {T.}~\bibnamefont {Ozawa}}, \bibinfo {author} {\bibfnamefont
  {N.}~\bibnamefont {Goldman}}, \ and\ \bibinfo {author} {\bibfnamefont
  {J.}~\bibnamefont {Cai}},\ }\href {\doibase 10.1093/nsr/nwz193} {\bibfield
  {journal} {\bibinfo  {journal} {National Science Review}\ }\textbf {\bibinfo
  {volume} {7}},\ \bibinfo {pages} {254} (\bibinfo {year} {2019})}\BibitemShut
  {NoStop}%
\bibitem [{\citenamefont {Yang}\ \emph {et~al.}(2019)\citenamefont {Yang},
  \citenamefont {Chiribella},\ and\ \citenamefont {Hayashi}}]{YangCMP2019}%
  \BibitemOpen
  \bibfield  {author} {\bibinfo {author} {\bibfnamefont {Y.}~\bibnamefont
  {Yang}}, \bibinfo {author} {\bibfnamefont {G.}~\bibnamefont {Chiribella}}, \
  and\ \bibinfo {author} {\bibfnamefont {M.}~\bibnamefont {Hayashi}},\ }\href
  {\doibase 10.1007/s00220-019-03433-4} {\bibfield  {journal} {\bibinfo
  {journal} {Communications in Mathematical Physics}\ }\textbf {\bibinfo
  {volume} {368}},\ \bibinfo {pages} {223} (\bibinfo {year}
  {2019})}\BibitemShut {NoStop}%
\bibitem [{\citenamefont {Suzuki}(2020)}]{SuzukiJPA2020}%
  \BibitemOpen
  \bibfield  {author} {\bibinfo {author} {\bibfnamefont {J.}~\bibnamefont
  {Suzuki}},\ }\href {\doibase 10.1088/1751-8121/ab8672} {\bibfield  {journal}
  {\bibinfo  {journal} {Journal of Physics A: Mathematical and Theoretical}\
  }\textbf {\bibinfo {volume} {53}},\ \bibinfo {pages} {264001} (\bibinfo
  {year} {2020})}\BibitemShut {NoStop}%
\bibitem [{\citenamefont {Suzuki}\ \emph {et~al.}(2020)\citenamefont {Suzuki},
  \citenamefont {Yang},\ and\ \citenamefont {Hayashi}}]{Suzuki_2020}%
  \BibitemOpen
  \bibfield  {author} {\bibinfo {author} {\bibfnamefont {J.}~\bibnamefont
  {Suzuki}}, \bibinfo {author} {\bibfnamefont {Y.}~\bibnamefont {Yang}}, \ and\
  \bibinfo {author} {\bibfnamefont {M.}~\bibnamefont {Hayashi}},\ }\href
  {\doibase 10.1088/1751-8121/ab8b78} {\bibfield  {journal} {\bibinfo
  {journal} {Journal of Physics A: Mathematical and Theoretical}\ }\textbf
  {\bibinfo {volume} {53}},\ \bibinfo {pages} {453001} (\bibinfo {year}
  {2020})}\BibitemShut {NoStop}%
\bibitem [{\citenamefont {Yang}(1978)}]{Yang1978}%
  \BibitemOpen
  \bibfield  {author} {\bibinfo {author} {\bibfnamefont {C.~N.}\ \bibnamefont
  {Yang}},\ }\bibfield  {booktitle} {\emph {\bibinfo {booktitle} {Journal of
  Mathematical Physics}},\ }\href {\doibase 10.1063/1.523506} {\bibfield
  {journal} {\bibinfo  {journal} {Journal of Mathematical Physics}\ }\textbf
  {\bibinfo {volume} {19}},\ \bibinfo {pages} {320} (\bibinfo {year}
  {1978})}\BibitemShut {NoStop}%
\bibitem [{\citenamefont {Hasebe}(2014)}]{Hasebe2014}%
  \BibitemOpen
  \bibfield  {author} {\bibinfo {author} {\bibfnamefont {K.}~\bibnamefont
  {Hasebe}},\ }\href {\doibase https://doi.org/10.1016/j.nuclphysb.2014.07.011}
  {\bibfield  {journal} {\bibinfo  {journal} {Nuclear Physics B}\ }\textbf
  {\bibinfo {volume} {886}},\ \bibinfo {pages} {952} (\bibinfo {year}
  {2014})}\BibitemShut {NoStop}%
\bibitem [{\citenamefont {Palumbo}\ and\ \citenamefont
  {Goldman}(2019)}]{PhysRevB.99.045154}%
  \BibitemOpen
  \bibfield  {author} {\bibinfo {author} {\bibfnamefont {G.}~\bibnamefont
  {Palumbo}}\ and\ \bibinfo {author} {\bibfnamefont {N.}~\bibnamefont
  {Goldman}},\ }\href {\doibase 10.1103/PhysRevB.99.045154} {\bibfield
  {journal} {\bibinfo  {journal} {Phys. Rev. B}\ }\textbf {\bibinfo {volume}
  {99}},\ \bibinfo {pages} {045154} (\bibinfo {year} {2019})}\BibitemShut
  {NoStop}%
\bibitem [{\citenamefont {Yu}\ \emph {et~al.}(2020)\citenamefont {Yu},
  \citenamefont {Liu}, \citenamefont {Yang}, \citenamefont {Gong},
  \citenamefont {Cao}, \citenamefont {Zhang}, \citenamefont {Liu},
  \citenamefont {Heyl}, \citenamefont {Ozawa}, \citenamefont {Goldman},\ and\
  \citenamefont {Cai}}]{Yuarxiv2020}%
  \BibitemOpen
  \bibfield  {author} {\bibinfo {author} {\bibfnamefont {M.}~\bibnamefont
  {Yu}}, \bibinfo {author} {\bibfnamefont {Y.}~\bibnamefont {Liu}}, \bibinfo
  {author} {\bibfnamefont {P.}~\bibnamefont {Yang}}, \bibinfo {author}
  {\bibfnamefont {M.}~\bibnamefont {Gong}}, \bibinfo {author} {\bibfnamefont
  {Q.}~\bibnamefont {Cao}}, \bibinfo {author} {\bibfnamefont {S.}~\bibnamefont
  {Zhang}}, \bibinfo {author} {\bibfnamefont {H.}~\bibnamefont {Liu}}, \bibinfo
  {author} {\bibfnamefont {M.}~\bibnamefont {Heyl}}, \bibinfo {author}
  {\bibfnamefont {T.}~\bibnamefont {Ozawa}}, \bibinfo {author} {\bibfnamefont
  {N.}~\bibnamefont {Goldman}}, \ and\ \bibinfo {author} {\bibfnamefont
  {J.}~\bibnamefont {Cai}},\ }\href {https://arxiv.org/abs/2003.08373}
  {\bibfield  {journal} {\bibinfo  {journal} {arXiv:2003.08373}\ } (\bibinfo
  {year} {2020})}\BibitemShut {NoStop}%
\bibitem [{\citenamefont {Hou}\ \emph {et~al.}(2020)\citenamefont {Hou},
  \citenamefont {Zhang}, \citenamefont {Xiang}, \citenamefont {Li},
  \citenamefont {Guo}, \citenamefont {Chen}, \citenamefont {Liu},\ and\
  \citenamefont {Yuan}}]{HouPRL2020}%
  \BibitemOpen
  \bibfield  {author} {\bibinfo {author} {\bibfnamefont {Z.}~\bibnamefont
  {Hou}}, \bibinfo {author} {\bibfnamefont {Z.}~\bibnamefont {Zhang}}, \bibinfo
  {author} {\bibfnamefont {G.-Y.}\ \bibnamefont {Xiang}}, \bibinfo {author}
  {\bibfnamefont {C.-F.}\ \bibnamefont {Li}}, \bibinfo {author} {\bibfnamefont
  {G.-C.}\ \bibnamefont {Guo}}, \bibinfo {author} {\bibfnamefont
  {H.}~\bibnamefont {Chen}}, \bibinfo {author} {\bibfnamefont {L.}~\bibnamefont
  {Liu}}, \ and\ \bibinfo {author} {\bibfnamefont {H.}~\bibnamefont {Yuan}},\
  }\href {\doibase 10.1103/PhysRevLett.125.020501} {\bibfield  {journal}
  {\bibinfo  {journal} {Phys. Rev. Lett.}\ }\textbf {\bibinfo {volume} {125}},\
  \bibinfo {pages} {020501} (\bibinfo {year} {2020})}\BibitemShut {NoStop}%
\bibitem [{\citenamefont {Hou}\ \emph {et~al.}(2021)\citenamefont {Hou},
  \citenamefont {Tang}, \citenamefont {Chen}, \citenamefont {Yuan},
  \citenamefont {Xiang}, \citenamefont {Li},\ and\ \citenamefont
  {Guo}}]{HouSA2021}%
  \BibitemOpen
  \bibfield  {author} {\bibinfo {author} {\bibfnamefont {Z.}~\bibnamefont
  {Hou}}, \bibinfo {author} {\bibfnamefont {J.-F.}\ \bibnamefont {Tang}},
  \bibinfo {author} {\bibfnamefont {H.}~\bibnamefont {Chen}}, \bibinfo {author}
  {\bibfnamefont {H.}~\bibnamefont {Yuan}}, \bibinfo {author} {\bibfnamefont
  {G.-Y.}\ \bibnamefont {Xiang}}, \bibinfo {author} {\bibfnamefont {C.-F.}\
  \bibnamefont {Li}}, \ and\ \bibinfo {author} {\bibfnamefont {G.-C.}\
  \bibnamefont {Guo}},\ }\href {\doibase 10.1126/sciadv.abd2986} {\bibfield
  {journal} {\bibinfo  {journal} {Science Advances}\ }\textbf {\bibinfo
  {volume} {7}} (\bibinfo {year} {2021}),\ 10.1126/sciadv.abd2986}\BibitemShut
  {NoStop}%
\bibitem [{\citenamefont {Synge}(1971)}]{SyngeRSPA1971}%
  \BibitemOpen
  \bibfield  {author} {\bibinfo {author} {\bibfnamefont {J.~L.}\ \bibnamefont
  {Synge}},\ }\href {\doibase 10.1098/rspa.1971.0162} {\bibfield  {journal}
  {\bibinfo  {journal} {Proc. R. Soc. London. A. Mathematical and Physical
  Sciences}\ }\textbf {\bibinfo {volume} {325}},\ \bibinfo {pages} {151}
  (\bibinfo {year} {1971})}\BibitemShut {NoStop}%
\bibitem [{\citenamefont {Qin}\ \emph {et~al.}(2016)\citenamefont {Qin},
  \citenamefont {Fei},\ and\ \citenamefont {Li-Jost}}]{QinSR2016}%
  \BibitemOpen
  \bibfield  {author} {\bibinfo {author} {\bibfnamefont {H.-H.}\ \bibnamefont
  {Qin}}, \bibinfo {author} {\bibfnamefont {S.-M.}\ \bibnamefont {Fei}}, \ and\
  \bibinfo {author} {\bibfnamefont {X.}~\bibnamefont {Li-Jost}},\ }\href
  {\doibase 10.1038/srep31192} {\bibfield  {journal} {\bibinfo  {journal}
  {Scientific Reports}\ }\textbf {\bibinfo {volume} {6}},\ \bibinfo {pages}
  {31192} (\bibinfo {year} {2016})}\BibitemShut {NoStop}%
\bibitem [{\citenamefont {Dodonov}(2018)}]{DodonovPRA2018}%
  \BibitemOpen
  \bibfield  {author} {\bibinfo {author} {\bibfnamefont {V.~V.}\ \bibnamefont
  {Dodonov}},\ }\href {\doibase 10.1103/PhysRevA.97.022105} {\bibfield
  {journal} {\bibinfo  {journal} {Phys. Rev. A}\ }\textbf {\bibinfo {volume}
  {97}},\ \bibinfo {pages} {022105} (\bibinfo {year} {2018})}\BibitemShut
  {NoStop}%
\bibitem [{\citenamefont {Palumbo}\ and\ \citenamefont
  {Goldman}(2018)}]{GianPRL2018}%
  \BibitemOpen
  \bibfield  {author} {\bibinfo {author} {\bibfnamefont {G.}~\bibnamefont
  {Palumbo}}\ and\ \bibinfo {author} {\bibfnamefont {N.}~\bibnamefont
  {Goldman}},\ }\href {\doibase 10.1103/PhysRevLett.121.170401} {\bibfield
  {journal} {\bibinfo  {journal} {Phys. Rev. Lett.}\ }\textbf {\bibinfo
  {volume} {121}},\ \bibinfo {pages} {170401} (\bibinfo {year}
  {2018})}\BibitemShut {NoStop}%
\bibitem [{\citenamefont {Mera}\ \emph
  {et~al.}(2022{\natexlab{b}})\citenamefont {Mera}, \citenamefont {Zhang},\
  and\ \citenamefont {Goldman}}]{Mera22}%
  \BibitemOpen
  \bibfield  {author} {\bibinfo {author} {\bibfnamefont {B.}~\bibnamefont
  {Mera}}, \bibinfo {author} {\bibfnamefont {A.}~\bibnamefont {Zhang}}, \ and\
  \bibinfo {author} {\bibfnamefont {N.}~\bibnamefont {Goldman}},\ }\href
  {\doibase 10.21468/SciPostPhys.12.1.018} {\bibfield  {journal} {\bibinfo
  {journal} {SciPost Phys.}\ }\textbf {\bibinfo {volume} {12}},\ \bibinfo
  {pages} {18} (\bibinfo {year} {2022}{\natexlab{b}})}\BibitemShut {NoStop}%
\end{thebibliography}%

\end{document}